\documentclass[12pt]{article}
\usepackage{amssymb,amsmath}
\usepackage{epsfig,psfrag}
\usepackage{curves}
\usepackage{subfigure}

\catcode`\@=11
\def \tr {\mathop{\rm tr}\nolimits}

\def \e  {\mathop{\rm e}\nolimits}
\newcommand\lr[1]{{\left({#1}\right)}}

\def \qqquad {\qquad\quad}
\def \qqqquad {\qquad\qquad}

\newcommand{\F}{{\widehat F}}

\newcommand{\cO}{{\cal O}}

\newcommand{\cK}{{\cal K}}

\newcommand{\ep}{\epsilon}

\newcommand{\be}{\begin{equation}}
\newcommand{\ee}{\end{equation}}
\newcommand{\bea}{\begin{eqnarray}}
\newcommand{\eea}{\end{eqnarray}}
\newcommand{\eaa}{\end{eqnarray}}

\newcommand{\cN}{{\cal N}}

\newcommand{\p}[1]{(\ref{#1})}
\newcommand{\bt}[1]{{\bar t}}
\newcommand{\ts}{\textstyle}

\newcommand{\half}{{\ts \frac{1}{2}}}
\newcommand \vev [1] {\langle{#1}\rangle}

\newcommand{\ft}[2]{{\textstyle\frac{#1}{#2}}}

\def \qqquad {\qquad\quad}
\def \qqqquad {\qquad\qquad}

\newcommand{\lm}{\lambda}
\newcommand{\Gm}{\Gamma}

\catcode`\@=11
\def \tr {\mathop{\rm tr}\nolimits}

\def \e  {\mathop{\rm e}\nolimits}

\def\numberbysection{\@addtoreset{equation}{section}
                     \def\theequation{\thesection.\arabic{equation}}}

\numberbysection

\begin{document}

\thispagestyle{empty}

\null\vskip-43pt \hfill
\begin{minipage}[t]{50mm}
CERN-PH-TH/2012-053 \\
DCPT-12/11 \\
HU-EP-12/07\\
HU-MATH 2012-04\\
IPhT--T12/014 \\
LAPTH-011/12
\end{minipage}

\vskip2.2truecm
\begin{center}
\vskip 0.2truecm

 {\Large\bf
Five-loop Konishi in $\mathcal{N}=4$ SYM }
\vskip 0.5truecm

\vskip 1truecm
{\bf    Burkhard Eden$^{a}$, Paul Heslop$^{b}$, Gregory P. Korchemsky$^{c}$, \\[2mm] Vladimir A. Smirnov$^{d}$,
Emery Sokatchev$^{e,f,g}$ \\
}

\vskip 0.4truecm
$^{a}$ {\it Institut f\"ur Mathematik, Humboldt-Universit\"at zu Berlin,
\\
Rudower Chaussee 25, Johann von Neumann-Haus, 12489 Berlin}
 \\
  \vskip .2truecm
$^{b}$ {\it  Mathematics Department, Durham University,
Science Laboratories,
 \\
South Rd, Durham DH1 3LE,
United Kingdom \\
 \vskip .2truecm
$^{c}$ Institut de Physique Th\'eorique\,\footnote{Unit\'e de Recherche Associ\'ee au CNRS URA 2306},
CEA Saclay, 
91191 Gif-sur-Yvette Cedex, France\\
\vskip .2truecm $^{d}$ Skobeltsyn Institute of Nuclear Physics, Moscow State University, 
 119992 Moscow, Russia \\
\vskip .2truecm $^{e}$ Physics Department, Theory Unit, CERN, CH -1211, Geneva 23, Switzerland \\
\vskip .2truecm $^{f}$ Institut Universitaire de France,  103, bd Saint-Michel
F-75005 Paris, France \\
\vskip .2truecm $^{g}$ LAPTH\,\footnote[2]{Laboratoire d'Annecy-le-Vieux de Physique Th\'{e}orique, UMR 5108},   Universit\'{e} de Savoie, CNRS,
B.P. 110,  F-74941 Annecy-le-Vieux, France
                       } \\
\end{center}

\vskip -.2truecm

\centerline{\bf Abstract} 
\medskip
\noindent
We present a new method for computing the Konishi anomalous dimension in $\cN=4$ SYM at weak coupling. It does not rely on the conventional Feynman diagram technique
and is not restricted to the planar limit. It is based on the OPE analysis of the four-point correlation function of stress-tensor multiplets, which has been recently constructed
up to six loops. The Konishi operator gives the leading contribution to the singlet $SU(4)$ channel of this OPE. Its anomalous dimension
is the coefficient of  the leading single logarithmic singularity of the logarithm of the correlation function in the double short-distance limit,
in which the operator positions coincide pairwise. We  regularize the logarithm of the correlation function in this singular limit by a version of
dimensional regularization. At any loop level, the resulting singularity is a simple pole whose residue is determined by a finite two-point integral
with one loop less. This drastically simplifies the five-loop calculation of the Konishi anomalous dimension by reducing it to a set of known
four-loop two-point integrals and two unknown integrals which we evaluate analytically. We obtain an analytic result at five loops in the
planar limit and observe perfect agreement with the prediction
based on integrability in AdS/CFT.

\newpage

\thispagestyle{empty}

{\small \tableofcontents}

\newpage
\setcounter{page}{1}\setcounter{footnote}{0}



\section{Introduction}

It has been realized recently that the four-point correlation function of the so-called
stress-tensor multiplets in $\mathcal{N}=4$ super-Yang-Mills theory (SYM) has
a new symmetry \cite{Eden:2011we}. In combination with $\cN=4$ superconformal symmetry, it imposes strong constraints on the integrand of the loop correction to the correlation
function and leads to an iterative structure at weak coupling,
at any loop order and for a gauge group of  arbitrary rank \cite{Eden:2012tu}. The correlation function of four stress-tensor
multiplets plays a special role in  $\mathcal{N}=4$ SYM theory. In virtue of the operator
product expansion (OPE), its asymptotic behaviour at short distances contains information
about the anomalous dimensions of a large variety of Wilson operators and the
corresponding structure constants of the OPE. Moreover, if considered in the planar limit
and restricted to the light cone, it is  dual to the four-particle scattering amplitudes \cite{Eden:2010zz,Eden:2010ce}.

For more than ten years, this correlation function was not
known beyond two loops. The main difficulty in going to higher loops is due to the factorially
increasing number of contributing Feynman diagrams. In general, each individual diagram
respects neither gauge invariance nor conformal symmetry, but the symmetries are restored in the sum of all diagrams. This calls for developing a new approach that makes full use of $\cN=4$ superconformal symmetry and of the specific symmetry of the stress-tensor multiplet
mentioned above.
Such an approach has been proposed in two recent papers
\cite{Eden:2011we,Eden:2012tu}, where a new construction of the four-point
correlation function was carried out in $\mathcal{N}=4$ SYM theory for the gauge group
$SU(N_c)$ with arbitrary $N_c$. In the planar limit, for $N_c\to\infty$ and with the 't Hooft coupling
$a=g^2N_c/(4\pi^2)$ fixed, the integrand of the four-point correlation function was   determined
up to six loops and the non-planar $O(1/N_c^2)$ correction was  identified at four loops (up to
four arbitrary rational constants).~\footnote{It is worth mentioning that our approach is
not limited to six loops. Extending it to higher orders is just a question of  computer power.}

In this paper, we apply the results of Refs.~\cite{Eden:2011we,Eden:2012tu} to
perform the OPE analysis of the four-point correlation function of the stress-tensor
multiplets. As the main result of our analysis, we present a new method for computing the
Konishi anomalous dimension in $\mathcal{N}=4$ SYM theory for arbitrary gauge group
$SU(N_c)$. The Konishi operator is the simplest unprotected gauge invariant Wilson
operator  in  $\mathcal{N}=4$ SYM, whose scaling dimension receives anomalous
contribution to all loops. In  the OPE context, the distinguishing feature of the
Konishi operator is that it controls the leading asymptotic behaviour of the four-point
correlation function at loop level in the short-distance limit. In this manner, we
obtain an analytic result for the Konishi anomalous dimension at five loops in
planar $\mathcal{N}=4$ SYM theory and observe perfect agreement with the prediction
based on integrability in AdS/CFT \cite{Bajnok:2008bm,Bajnok:2009vm,Arutyunov:2010gb,Balog:2010xa}.

The properties of the Konishi operator have been studied extensively after the discovery of the so-called Konishi anomaly \cite{Konishi:1983hf} in supersymmetric gauge theories.  The interest in the subject was renewed in the context of the AdS/CFT correspondence. As was observed in \cite{Andrianopoli:1998jh}, the Konishi supermultiplet in $\cN=4$ SYM  theory is a long (or unprotected) multiplet that corresponds to the first string level in the spectrum of type IIB excitations in an AdS${}_5\times S^5$ background.
Recently, the Konishi anomalous dimension $\gamma_\cK(a)$ again attracted a lot of attention after the  discovery of integrability  in the planar limit on both sides of the AdS/CFT correspondence (for a review see \cite{Beisert:2010jr}). At strong
coupling, the first few terms of the expansion of $\gamma_\cK(a)$
in powers of $a^{-1/4}$ in  the planar limit were obtained from the semiclassical quantization of short
strings on an AdS${}_5\times S^5$ background \cite{Vallilo:2011fj,Roiban:2011fe,Gromov:2011de}. At weak coupling,  the values of $\gamma_\cK(a)$ at four and five loops in planar  $\cN=4$  SYM were predicted in Refs.~\cite{Bajnok:2008bm}
and \cite{Bajnok:2009vm}, respectively, from the integrable string
sigma model by evaluating finite size effects using L\"uscher's formulas.\footnote{See also
\cite{Arutyunov:2010gb,Balog:2010xa} for an alternative approach using the mirror thermodynamic Bethe
Ansatz.}  The four-loop prediction was later confirmed by direct
perturbative calculations using $\cN=1$  Feynman super-graphs
\cite{Fiamberti:2007rj,Fiamberti:2008sh} and ordinary Feynman diagrams \cite{Velizhanin:2008jd}. Until now, no five-loop test of the integrability
prediction had been performed, and it is not clear whether traditional
techniques would allow one to reach such a high perturbative level.
 A numerical prediction for  $\gamma_\cK(a)$ at intermediate coupling, interpolating between the strong and weak coupling results, was made in \cite{Gromov:2009zb} from the solution of the $Y-$system of integral non-linear equations and more recently in \cite{Frolov:2010wt} from the TBA equations.

As was already mentioned, the Konishi operator provides the leading contribution
to the asymptotic behaviour of the four-point correlation function at short distances
$G(1,2,3,4)\sim (x_{12}^2 )^{\gamma_\cK(a)/2}$ as $x_1\to x_2$ (with $x_{12} \equiv x_1-x_2$).
At weak coupling, this asymptotic behaviour implies that perturbative corrections to the correlation function at $\ell$ loops are given by a sum of logarithmic singularities $(\ln x_{12}^2)^k$ with powers $k\le \ell$. The coefficients of the higher powers of logarithms
(for $k>1$) are expressed in terms of the anomalous dimensions at lower loops. It is only
the single logarithm (with $k=1$) that carries information about the anomalous dimension
at $\ell$ loops. This fact complicates the evaluation of the anomalous dimension.
It is more advantageous to consider instead the logarithm of the correlation function
$\ln G(1,2,3,4)$, whose asymptotic behaviour at short distances involves a single logarithmic singularity to all loops. We can further simplify the analysis by considering the double short-distance limit $x_1\to x_2$ and $x_3\to x_4$, in which case
$\ln G(1,2,3,4)\sim(\gamma_\cK(a)/2)\left[ \ln x_{12}^2+ \ln x_{34}^2\right]$.
To determine the anomalous dimension in this way, we need an efficient way of computing the perturbative corrections to  $\ln G(1,2,3,4)$.

Applying the results of Ref.~\cite{Eden:2012tu}, we can express  $\ln G(1,2,3,4)$ at $\ell$ loops as a
Euclidean integral, whose integrand is a conformally covariant function of the four
external points and the $\ell$ integration points. In the short-distance limit $x_1\to x_2$,
the integral develops a single logarithmic singularity $\sim \ln x_{12}^2$ when all $\ell$
integration points approach the external point $x_1$ simultaneously. Similarly,
for $x_3\to x_4$ the logarithmic singularity $\sim \ln x_{34}^2$ originates from
integration in the vicinity of the point $x_3$. It is clear that these singularities
are of ultraviolet (UV) origin with the small distances $x_{12}^2$ and $x_{34}^2$ playing
the role of an UV cut-off. To extract the coefficient of the logarithmic singularity of the
integral, which defines the anomalous dimension of the Konishi operator, we can simplify
the calculation by  taking $x_{12}=x_{34}=0$ inside the integral and by introducing the
most convenient regularization scheme for the resulting UV divergences. This is done by changing the integration measure from
$D=4$ to $D=4-2\ep$ dimensions.
Thus, we transform the expected single logarithmic singularity of $\ln G(1,2,3,4)$
in the double short-distance limit into a simple pole $1/\ep$.
Our final simplification comes from the observation that, for $x_1=x_2$ and $x_3=x_4$, the $\ell-$loop residue at this
simple pole is in fact given by an $(\ell-1)-$loop {\it finite} two-point integral of the propagator type.
We can then apply an array of well-known and very efficient methods for computing such integrals.

In summary, we have reduced the problem of computing the Konishi anomalous
dimension at $\ell$ loops to the problem of evaluating a {\it finite two-point integral at
$(\ell-1)$ loops}.\footnote{This feature is typical for usual renormalization group calculations in momentum space \cite{RStar2}.}
This allowed us to obtain the following result up to five loops
\footnote{The one-loop value  of $\gamma_\cK$ was found for the first
  time in \cite{Anselmi:1996mq}. At two loops, it was first extracted from the OPE of two stress-tensor multiplets in  $\cN=4$ SYM, as part of the investigation of the two-loop four-point correlation functions of  half-BPS operators \cite{Eden:2000mv,Bianchi:2000hn}.
The three-loop value, together with the anomalous dimensions of all
twist-two Wilson operators in $\cN=4$ SYM was originally conjectured
in \cite{Kotikov:2004er}. This three-loop prediction for $\gamma_\cK$
was then confirmed for the first time in \cite{Eden:2004ua} by a direct   perturbative calculation. }:
\begin{align}\label{Main} \notag
\gamma_\mathcal{K}(a) = 3 \, a - 3\, a^2  & + \frac{21}4\,a^3 - \lr{\frac{39}4 - \frac94 \zeta_3 + \frac{45}8 \zeta_5 -\frac{r}{N_c^2} \zeta_5}a^4
\\
&
 +
\lr{\frac{237}{16}+\frac{27}{4} \zeta_3-\frac{81}{16} \zeta _3{}^2-\frac{135}{16} \zeta_ 5+\frac{945}{32}\zeta_7+ O(1/N_c^2)}a^5 + O(a^6)\,,
\end{align}
where
 the non-planar four-loop correction is predicted  up to an arbitrary
 {rational} constant, $r$.
We use $\zeta_n$ to denote values of the Riemann zeta-function $\zeta(n)$ at integer points.
The obtained expression for $\gamma_\mathcal{K}(a)$ is in agreement
with the existing perturbative four-loop results of \cite{Fiamberti:2007rj,Fiamberti:2008sh,Velizhanin:2008jd,Velizhanin:2009gv}
and with the five-loop prediction of \cite{Bajnok:2008bm,Bajnok:2009vm,Arutyunov:2010gb,Balog:2010xa} based on integrability in AdS/CFT.
As a byproduct of the OPE analysis of the four-point correlation function, we also investigated the spectrum of anomalous dimensions of the twist-two operators with non-vanishing spin at three loops
and found agreement with the values conjectured in \cite{Kotikov:2004er}.

The paper is organized as follows. In Section 2 we define the four-point correlation function
of stress-tensor multiplets in $\mathcal{N}=4$ SYM and use the OPE to relate  $\gamma_\cK(a)$ to the leading asymptotic behaviour of the correlation function in the
short distance limit. In Section 3 we formulate our method for computing the Konishi anomalous dimension and illustrate it by evaluating $\gamma_\mathcal{K}(a)$ up to
two loops.  In Section 4 we extend our analysis to four loops. Applying the results
of Refs.~\cite{Eden:2012tu} for the four-loop correlation function and making use of well-known
techniques for evaluating Feynman integrals, we express the four-loop correction to $\gamma_\mathcal{K}(a)$ as a linear combination of six master two-point three-loop integrals (five in the planar sector and only one in the non-planar sector). If rewritten in the dual
momentum representation, the latter coincide with some known
finite three-loop integrals of the  propagator type. In Section 5 we evaluate the five-loop correction
to  $\gamma_\cK(a)$ in the planar limit. We show that it is given
by a linear combination of 22 master scalar four-loop integrals. Among them 20 integrals
correspond to planar graphs and coincide, in the dual momentum
representation, with known finite four-loop propagator integrals~\cite{BC}.
The remaining two non-planar integrals are evaluated in Appendix~\ref{app-master}.
Section 6 contains concluding remarks. In Appendix A, we describe the method of
IR rearrangement in the configuration space that we employ in our calculation of  $\gamma_\cK(a)$. In Appendix C, we perform the OPE analysis of the
four-point correlation function and extract the values of three-loop anomalous dimensions of twist-two operators with Lorentz spin zero, two and four.

\section{Four-point correlation function}

\subsection{Expression for the integrand}

In this paper we   study the OPE of the stress-tensor multiplet  in $\mathcal{N}=4$ SYM. This is the simplest example of a
half-BPS operator, whose superconformal primary state has the form
\begin{align}\label{O-IJ}
  \mathcal{O}_{\mathbf{20'}}^{IJ} &= \tr\left(\Phi^I \Phi^J\right) - \frac16 \delta^{IJ} \tr\left(\Phi^K \Phi^K\right)\,.
\end{align}
It is built from the six real scalars $\Phi^I$ (with $I=1,\ldots,6$ being an $SO(6)$ index) in the adjoint representation of the gauge group $SU(N_c)$ and belongs to the representation
$\mathbf{20'}$ of the $R$ symmetry group $SO(6)\sim SU(4)$. To keep track of the $SO(6)$ tensor structure of the OPE, it proves convenient to introduce auxiliary $SO(6)$ harmonic variables $Y_I$, defined as a (complex)
null vector, $Y^2\equiv Y_I Y_I=0$, and project the indices of  $\mathcal{O}^{IJ}$ as follows:
\begin{align}\label{defO}
{\mathcal O}(x,y) \equiv Y_I \, Y_J\, \mathcal{O}_{\mathbf{20'}}^{IJ}(x) =Y_I \, Y_J\,  \tr\left(\Phi^I(x) \Phi^J(x)\right)  \,,
\end{align}
where $y$ denotes the dependence on the $Y-$variables.

An important property of the operator \p{O-IJ} is that its scaling dimension is protected
from perturbative corrections. The same is true for the two- and three-point correlation
functions of the operator $\mathcal{O}(x_i,y_i)$. The four-point   correlation function is the first to receive perturbative corrections:
\begin{equation}
G_4=\vev{ {\mathcal O}(x_1,y_1) {\mathcal O}(x_2,y_2) {\mathcal O}(x_3,y_3)    {\mathcal O}(x_4,y_4)}   =   \sum_{\ell=0}^\infty a^{\ell}
 \, G^{(\ell)}_{4}(1,2,3,4)\,,
\label{cor4loop}
\end{equation}
where the expansion on the right-hand side runs in powers of the 't Hooft coupling $a=g^2N_c/(4\pi^2)$ and $G^{(\ell)}_{4}$  denotes the perturbative correction at $\ell$ loops.
Notice that here we do not assume the planar limit and allow $G^{(\ell)}_{4}$ to have a  non-trivial dependence  on $N_c$. It is this four-point function that will serve as the starting point of our OPE analysis.

At tree level, $G^{(0)}_{4}$ reduces to a product of free scalar propagators and the
corresponding expression can be found in Ref.~\cite{Eden:2011we}. At loop level, the superconformal
symmetry of the $\mathcal{N}=4$ SYM theory restricts $G^{(\ell)}_{4}$ to have  the following factorized form  \cite{Eden:2000bk,Eden:2011we}:
\begin{align}\label{intriLoops}
  G_{4}^{(\ell)}(1,2,3,4)= \frac{2 \, (N_c^2-1)}{(4\pi^2)^{4}} \times R(1,2,3,4)   \times  F^{(\ell)}(x_i)  \qquad \mbox{(for $\ell \ge 1$)} \,,
\end{align}
where $F^{(\ell)}(x_i)$ is a function of $x_i$ only (with $i=1,2,3,4$) to be specified below
and  $R(1,2,3,4)$ is a universal, $\ell-$independent
rational function of the space-time, $x_i$, and harmonic, $Y_I$, coordinates at the four external points $1,2,3,4$, whose explicit  expression can be found in Ref.~\cite{Eden:2011we}.

So from \p{cor4loop} and \p{intriLoops} we find that the loop corrections to the four-point correlation function are determined by a single function $F^{(\ell)}(x_i)$. As was shown
in Refs.~\cite{Eden:2011we,Eden:2012tu}, this function has a number of remarkable properties in $\mathcal{N}=4$ SYM theory. Namely, it can be represented in the form  of an $\ell-$loop Euclidean integral,
\begin{align}\label{integg}
  F^{(\ell)}(x_i)={x_{12}^2 x_{13}^2 x_{14}^2 x_{23}^2
  x_{24}^2 x_{34}^2\over \ell!\,(-4\pi^2)^\ell} \int d^4x_5 \dots
  d^4x_{4+\ell} \,  f^{(\ell)}(x_1,  \dots , x_{4+\ell})\,,
\end{align}
where the integrand $f^{(\ell)}$  depends on the four external coordinates $x_1,\ldots,x_4$
and the $\ell$ additional (internal) coordinates $x_5,\ldots,x_{4+\ell}$ giving the positions of  the Lagrangian insertions. The integrand $f^{(\ell)}$ can be written in the form
\begin{align}
   \label{eq:10}
   f^{(\ell)}(x_1, \dots, x_{4+\ell})= { P^{(\ell)}(x_1, \dots ,
   x_{4+\ell}) \over \prod_{1\leq i<j \leq 4+\ell}   x_{ij}^2}\ .
 \end{align}
Here the denominator contains the product of all distances between the $(4+\ell)$ points
and $P^{(\ell)}$ is a homogeneous polynomial in $x_{ij}^2$ of degree $(\ell-1)(\ell+4)/2$.
Most importantly, this polynomial is  symmetric under the exchange of any pair of points
$x_i$ and $x_j$ (both external and internal). As we have shown in Refs.~\cite{Eden:2012tu}, this
property alone combined with the correct asymptotic behaviour of the correlation function
in the short-distance and the light-cone limits, allows us to completely determine $F^{(\ell)}(x_i)$ up to six loops in the planar sector. In the non-planar sector, the same analysis leads
to an expression depending on a few constants only. For example, at one and two loops,
we have
\begin{align}  \label{P2}
& P^{(1)} = 1\,,\qquad P^{(2)} =  \frac1{48} \sum_{\sigma\in S_6} x_{\sigma_1\sigma_2}^2x_{\sigma_3\sigma_4}^2x_{\sigma_5\sigma_6}^2 = x_{12}^2x_{34}^2x_{56}^2 + \dots \,,
\end{align}
where in the second relation the sum runs over all $S_6-$permutations of the indices $1,\ldots,6$.
Similar expressions at higher loops can be found in Ref.~\cite{Eden:2012tu}.

\subsection{Operator product expansion}\label{sect:OPE}

As was mentioned in the previous subsection, the  four-point
correlation function \p{integg} has a particular asymptotic behaviour at short
distances, dictated by the operator product expansion (OPE).

For the scalar operators  \p{defO} the OPE takes
the following form
\begin{align}\label{OPE0}
 {\mathcal O}(x_1,y_1) {\mathcal O}(x_2,y_2) = c_{\mathcal{I}} \frac{(Y_1\cdot Y_2)^2}{x_{12}^4}
 \mathcal{I} + c_{\mathcal{K}}(a) \frac{(Y_1\cdot Y_2)^2}{(x_{12}^2)^{1-\gamma_{\mathcal{K}}/2}} \mathcal{K}(x_2) + c_{\mathcal{O}} \frac{(Y_1\cdot Y_2)}{x_{12}^2} Y_{1I} Y_{2J} \mathcal{O}_{\mathbf{20'}}^{IJ}(x_2)+\ldots
\end{align}
where we only displayed the contribution of operators with naive scaling dimension up to two.
Here the most singular $1/x_{12}^4$ contribution comes from the identity operator $\mathcal{I}$
while the first subleading $O(1/x_{12}^2)$ contribution originates from two operators: the
half-BPS operator \p{O-IJ} and the Konishi operator defined as
\begin{align}\label{Konishi}
\mathcal{K} = \tr\left(\Phi^I\Phi^I\right)  \,.
\end{align}
Since the operators $\mathcal{I}$ and $\mathcal{O}_{\mathbf{20'}}$ are protected, the constants $c_{\mathcal{I}}$ and $c_{\mathcal{O}}$ do not depend on the coupling constant
and keep their tree-level values,  $c_{\mathcal{I}}=(N_c^2-1)/(32\pi^4)$ and $c_{\mathcal{O}}=1/(2\pi^2)$. For the Konishi operator, both the coefficient $c_{\mathcal{K}}(a)$ and
its scaling dimension $\Delta_\mathcal{K}(a)$ receive perturbative corrections to all loops.
In what follows we will mostly concentrate on the anomalous dimension of the Konishi operator~\footnote{The Konishi operator is the simplest of an infinite series of twist-two operators contained in the OPE \p{OPE0}. In Appendix C we extract from the OPE  the three-loop anomalous dimensions of the twist-two operators with Lorentz spin two and four.}
\begin{align}\label{K-anom}
 & \Delta_\mathcal{K} \, = \, 2 +
\gamma_\mathcal{K}(a) = 2 + \sum_{\ell=1}^\infty a^\ell \gamma_\mathcal{K}^{(\ell)}\,.
\end{align}

In the singular limit $x_1\to x_2$ we can apply the OPE expansion \p{OPE0} to find the asymptotic behaviour of the correlation function \p{intriLoops} at short distances (in Euclidean kinematics). It receives contributions from all operators on the right-hand side of  \p{OPE0}. A crucial advantage of the Konishi operator is that it has the minimal possible scaling dimension among all unprotected operators. To separate the contribution of the Konishi operator it is useful
to consider the double short-distance limit $x_1\to x_2$, $x_3\to x_4$.
 Taking into account the  relation  \p{OPE0} we  obtain the
asymptotic behaviour of the four-point correlation function in this limit as \cite{Eden:2011we}
\begin{align} \label{expect} \notag
G_4   \
\stackrel{x_2\to x_1 \atop x_4\to x_3}{\longrightarrow} \  \frac{(N_c^2-1)^2}{4(4\pi^2)^4}\frac{y_{12}^4 y_{34}^4}{x_{12}^4 x_{34}^4}
+\frac{N_c^2-1}{(4\pi^2)^4}\bigg[&  \frac{y_{12}^2 y_{34}^2 ( y_{13}^2 y_{24}^2 + y_{14}^2 y_{23}^2 )}{x_{12}^2 x_{34}^2 x_{13}^4}
\\
&    +\frac13  \frac{y_{12}^4 y_{34}^4}{x_{12}^2 x_{34}^2 x_{13}^4} \, \left( c_\cK^2(a)
u^{ {\gamma_\cK(a)}/{2}} -1  \right)  \bigg] \, + \, \ldots\ ,
\end{align}
where  $u$ is a conformal cross-ratio defined in \p{cr} below, $y_{ij}^2=(Y_i\cdot Y_j)$ denotes the scalar product of harmonic variables and the dots denote subleading terms.

Comparing the OPE prediction \p{expect} with the general expression for the correlation function, Eqs.~\p{cor4loop} and \p{intriLoops}, we obtain the following relation for the functions $F^{(\ell)}(x_i)$ for $x_2\to x_1$ and $x_4\to x_3$
 \begin{align}\label{cons}
  \sum_{\ell\ge 1} a^\ell F^{(\ell)}(x_i)  \
\stackrel{x_2\to x_1 \atop x_4\to x_3}{\longrightarrow} \
\frac1{6x_{13}^4}
\left(c_{\mathcal K}^2(a)
u^{{\gamma_{\mathcal K}(a)}/{2}} -1 \right) \times \big[1+ O(u)+O(1-v)\big]  \,.
\end{align}
Here $u$ and $v$ are the two conformally invariant cross-ratios made of the four points $x_i$,
\begin{align}\label{cr}
u=\frac{x^2_{12}x^2_{34}}{x^2_{13}x^2_{24}}\,, \qqquad
v=\frac{x^2_{14}x^2_{23}}{x^2_{13}x^2_{24}} \,,
\end{align}
so that $u\to 0$, $v \to 1$ in the double short-distance limit  $x_2\to x_1\,, x_4\to x_3$.
For our purposes it is convenient to introduce the notation for the function $ x_{13}^4 F^{(\ell)}(x_i)$ in this
limit,
\begin{align}\label{F-hat}
 x_{13}^4 F^{(\ell)}(x_i)  \
\stackrel{x_2\to x_1 \atop x_4\to x_3}{\longrightarrow} \  \F(x_i) \,,
\end{align}
and to rewrite the OPE limit \p{cons} as
\begin{align}\label{SD}
 \ln  \lr{1+6\sum_{\ell\ge 1} a^\ell \F^{(\ell)}(x_i)} \
\stackrel{u\to 0 \atop v\to 1}{\longrightarrow} \  \frac12
\gamma_{\mathcal K}(a) \ln u+ \ln \left(c_{\mathcal K}^2(a)\right) +  O(u)+ O((1-v))\,.
\end{align}

Let us now expand both sides of the relations \p{cons} and \p{SD} in the powers of the coupling $a$
and compare their short-distance asymptotics. We find from \p{cons} that
$\F^{(\ell)}(x_i)  \sim (\ln u)^\ell$ as $u\to 0$. In particular,   from  \p{cons} we have to two-loop order
\begin{align}\notag
\F^{(1)} &=\frac1{12}\gamma^{(1)}_\cK \ln u  + \frac12 \alpha^{(1)} + \dots\,,
\\[2mm]  \label{ex}
\F^{(2)} &=\frac1{48} (\gamma^{(1)}_\cK)^2\, (\ln u)^2+\left(\frac1{12}\gamma^{(2)}_\cK+\frac14 \gamma^{(1)}_\cK\alpha^{(1)}
\right)\ln u  + \frac12 \alpha^{(2)} + \dots\,,
\end{align}
where the constants $\gamma^{(\ell)}_\cK$ and $\alpha^{(\ell)}$
define the perturbative corrections to the anomalous dimension $\gamma_\mathcal{K}(a)=\sum_{\ell \ge 1} a^\ell \gamma^{(\ell)}_\cK$
and to the coefficients $(c_{\mathcal{K}}(a))^2=1+3\sum_{\ell \ge 1}
a^\ell \alpha^{(\ell)}$. In a similar manner,  from \p{SD} we obtain
that the particular combination of the functions $\F^{(\ell)}(x_i)$  with $1\le \ell'\le \ell$,
arising from the expansion of the logarithm on the left-hand side of \p{SD}, scales
as $\ln u$. For instance,  at two-loop order we have from \p{SD}
\begin{align}\label{ex1}
\F^{(2)} - 3\, (\F^{(1)})^2 =  \frac1{12}\gamma^{(2)}_\cK \ln u  + \frac12 \alpha^{(2)} - \frac34 (\alpha^{(1)})^2 + \dots\,.
\end{align}
Comparing this relation with \p{ex}, we observe that the two-loop correction to the anomalous
dimension $\gamma^{(2)}_\cK$ appears in \p{ex} in the subleading term, while in \p{ex} it defines
the leading singular behaviour.  As we show in the next section, this property can be used to
drastically simplify the calculation of the Konishi anomalous dimension $\gamma_{\mathcal K}(a)$.

\section{Method for computing the Konishi anomalous\\ dimension}

Here we present our method for computing the Konishi anomalous dimension at higher
loops. It takes full advantage of the known properties of the correlation function explained in the
previous section. In this section we illustrate the key features of the method with the simplest examples of one and two loops.

Before we do this, we would like to recall the standard approach for extracting  the Konishi anomalous dimension from the asymptotic logarithmic behaviour of the four-point correlation function (see, e.g., Refs.~\cite{Bianchi:1999ge,Bianchi:2000hn}).   With the help of the relations \p{integg} -- \p{P2} we obtain the following expressions
for the correlation function to two loops:
 \begin{align}\label{F12}
F^{(1)} & =  g(1,2,3,4)\,, \notag
\\[2mm] \notag
 F^{(2)} & = h(1,2;3,4) + h(3,4;1,2) + h(2,3;1,4) +  h(1,4;2,3)
 \\ & + h(1,3;2,4) + h(2,4;1,3)
  +
 \frac12 \lr{x_{12}^2x_{34}^2+ x_{13}^2 x_{24}^2+x_{14}^2x_{23}^2}  [g(1,2,3,4)]^2\,.
\end{align}
Here the notation was introduced for the one- and two-loop conformal Euclidean integrals
\begin{align}
\notag
& g(1,2,3,4)  = - \frac{1}{4 \pi^2}
\int \frac{d^4x_5}{x_{15}^2 x_{25}^2 x_{35}^2 x_{45}^2}  \, , \\
\label{eq:g+h}
& h(1,2;3,4)  =  \frac{x^2_{34}}{(4 \pi^2)^2}
\int \frac{d^4x_5 \, d^4x_6}{(x_{15}^2 x_{35}^2 x_{45}^2) x_{56}^2
(x_{26}^2 x_{36}^2 x_{46}^2)}  \,,
\end{align}
with the remaining $h-$integrals obtained by permuting the indices.

The explicit expressions for these integrals as functions of the conformal ratios \p{cr} are known \cite{davussladder},
but what we need here is just their asymptotic behaviour
for $x_1\to x_2$ and $x_3\to x_4$, or equivalently
$u\to 0$ and $v\to 1$.  Replacing the integrals in \p{F12} by their asymptotic expansions, we easily obtain the following
result for the one- and two-loop correlation functions in the singular short-distance limit:
\begin{align} \label{F2}   \nonumber
\F^{(1)} & =   \frac{1}{4} \, \ln u  - \frac12  +\ldots  \, , \qquad
  \\
\F^{(2)} &=   \frac{3}{16} \, (\ln u)^2  -   \ln u +
\frac{3}{4} \, \zeta_3 + \frac{7}{4}  +\ldots  \, ,
\end{align}
leading to
\begin{align}\label{spec}
\F^{(2)} - 3\, (\F^{(1)})^2 = -\frac14 \ln u +\frac34\zeta_3+1+\ldots\ .
\end{align}
These relations are in perfect agreement with the OPE
prediction \p{ex} and \p{ex1}. They allow us to reproduce the well-known result
for the two-loop Konishi anomalous dimension, Eq.~\p{Main},
and  the two-loop normalization coefficients, $ \alpha^{(1)}=-1$ and  $\alpha^{(2)}=3\zeta_3/2+7/2$.

\subsection{One loop}

Let us now return to the one-loop expression $\F^{(1)}$, Eq.~\p{F2}, and understand the origin
of the singularity $\F^{(1)}\sim \ln u$ at short distances. It is easy to see from \p{eq:g+h} that for
$x_1\to x_2$  and $x_3\to x_4$ the integral $g(1,2,3,4) $ develops a logarithmic divergence coming
from the two distinct integration regions $x_5\to x_1$ and $x_5\to x_3$,
\begin{align}\label{delta}
\F^{(1)}  \sim -\frac1{4\pi^2} \int_{x_{51}^2 < \delta^2} \frac{d^4 x_5}{x_{15}^2x_{25}^2}   -\frac1{4\pi^2} \int_{x_{53}^2<  \delta^2} \frac{d^4 x_5}{x_{35}^2x_{45}^2 }   \,.
\end{align}
Here we have  restricted the integration to two balls of radius $\delta$, centred at the points $x_1 \sim x_2$ and $x_3 \sim x_4$. Choosing $ x^2_{13} \gg \delta^2 \gg x_{12}^2, x_{34}^2$ allows us to replace the other two propagator factors in the first integral  $x^2_{35} \sim x^2_{45}$ by $x^2_{13}$, and similarly for the second. This simplification of the integrand by replacing it with its asymptotic expression in the relevant integration region will be very helpful at higher loops.   Going to radial coordinates we find
\begin{align}\label{F-delta}
\F^{(1)}  \sim -\frac1{4} \int^{\delta^2}_{x_{12}^2} \frac{d x_{51}^2}{x_{51}^2}  -\frac1{4} \int^{\delta^2}_{x_{34}^2} \frac{d x_{53}^2}{x_{53}^2} =  \frac14 \ln \lr{x_{12}^2x_{34}^2/\delta^4} +\ldots\,,
\end{align}
where the short distances $x_{12}^2, x_{34}^2 \to 0$ serve as UV cut-offs and the dots denote terms  finite in the limit  $x_{12}^2, x_{34}^2\to 0$. It is easy to see that this relation is in agreement
with the first relation in \p{F2}.

Let us now examine what happens if we interchange  the integration in \p{delta} with taking the limit $x_1\to x_2$, $x_3\to x_4$. In this limit, the first integral on the right-hand side of \p{delta}
reduces to $\int_{x_{51}^2 < \delta^2}  d^4 x_5/x_{51}^4$ and it diverges as $x_5\to x_1$.
This is not surprising since $x_{12}^2$ plays the role of a short-distance cut-off in the first integral in \p{delta} and \p{F-delta}. Therefore, in order
to define the integral for $x_{12}^2=0$ we have to introduce a different short-distance regulator. The simplest way to do this is
to modify the integration measure in \p{delta} as follows
\footnote{We would like to emphasize that this regularization is different from the
conventional dimensional regularization in coordinate space in the sense that we modify the integration measure only and
use the scalar propagators $1/x^2$ instead of $1/(x^2)^{1-\epsilon}$. This explains why $\epsilon$ should be kept negative.}
\begin{align}\label{scheme}
d^4 x_5 \to \mu^{2\epsilon}\,d^{4-2\epsilon} x_5 \,,\qquad \text{(with $\epsilon<0$)}\,,
\end{align}
without changing the form of the integrand. In this way, we find from \p{delta}
\begin{align}\label{F-eps}
\F_\epsilon^{(1)}  \sim -\frac{\mu^{2\epsilon}}{4\pi^2} \int_{x_{51}^2 < \delta^2} \frac{d^{4-2\epsilon}  x_5}{(x_{51}^2)^2}   -\frac{\mu^{2\epsilon}}{4\pi^2} \int_{x_{53}^2<  \delta^2} \frac{d^{4-2\epsilon}  x_5}{(x_{53}^2)^2 }\,,
\end{align}
where we introduced the subscript in $\F_\epsilon^{(1)}$ to indicate that it is defined in the regularization
scheme \p{scheme} at $x_{12}^2=x_{34}^2=0$. Going to spherical coordinates,
$d^{4-2\epsilon} x =S_\epsilon \, r^{3-2\epsilon} dr$ with $S_\epsilon= 2\pi^{2-\epsilon}/\Gamma(2-\epsilon)$, we find that
the two integrals in \p{F-eps} produce equal contributions, leading to
\begin{align}\label{pole}
\F_\epsilon^{(1)}  = \frac{(\delta^2/\mu^2)^{-\ep}}{2\epsilon}  + O(\ep^0)= \frac1{2\epsilon} + \frac12 \ln(\mu^2/\delta^2) + \dots \,.
\end{align}
Comparing the right-hand sides of the relations \p{F-delta} and \p{pole}, we observe that they coincide
(up to the $O(1/\epsilon)$ term) upon the identification $x_{12}^2\to \mu^2$ and $x_{34}^2\to\mu^2$. In other words, for $x_{12}^2=x_{34}^2=0$,
within the regularization scheme \p{scheme}, the dimensionful parameter $\mu^2$ plays the role of the UV cut-off.
This property allows us to relate the coefficient in front of $\ln u$ in the asymptotic behaviour of $F^{(1)}  $ at small $u$, Eq.~\p{F2}, to the residue of $F_\epsilon^{(1)}$ at the simple pole  $1/\epsilon$. Moreover, it follows from \p{ex} that this coefficient coincides with
the one-loop Konishi anomalous dimension  $\gamma_\mathcal{K}^{(1)}$, leading to
\begin{align}\label{gamma1}
\gamma_\mathcal{K}^{(1)}=12\frac{d}{d\ln u} \F^{(1)} =  6\frac{d}{d\ln \mu^2}  \F_\epsilon^{(1)}  = 3\,,
\end{align}
in agreement with \p{Main}.

This suggests a new method for computing the Konishi anomalous dimension: Instead of evaluating the finite
four-dimensional integrals in \p{F12} and finding their asymptotics at $u\to0, v \to 1$ afterwards, we can first
evaluate the integrand at $x_1=x_2$ and $x_3=x_4$, thus making the integrals divergent, then introduce
the regularization \p{scheme} and, finally, identify the terms singular for $\epsilon\to 0$.

We would like to emphasize that this method captures correctly only the terms divergent for $u\to0, v \to 1$ but not
the finite ones. To see this, let us apply the above procedure to the one-loop expression $\F^{(1)}$
\begin{align}\label{F-F}
\F^{(1)}\  \stackrel{x_{12}, x_{34}= 0}{\longrightarrow} \ \F^{(1)}_\epsilon =- \frac{\mu^{2\epsilon}}{4 \pi^2}
\int \frac{d^{4-2\epsilon} x_5\,  x_{13}^4}{x_{15}^4 x_{35}^4}\,.
\end{align}
In comparison with \p{F-eps} here we did not restrict the integration region over $x_5$.  This is not really necessary, since the integral converges at large $x_5$.
To perform the integration, it is convenient to switch to the dual momenta $k=x_{15}$ and
$p=x_{13}$. Then, the integral in \p{F-F} takes the form of the standard one-loop ``bubble'' momentum integral of propagator type:
\begin{align}\label{I1}
M^{(1)}=-\frac{\mu^{2\epsilon}}{4 \pi^2}
\int \frac{ d^{4-2\epsilon} k\, p^4}{k^4 (p-k)^4}\,,
\end{align}
leading to (with $\bar\mu^2 = \mu^2/(\e^{\gamma_{\rm E}}\pi)$)
\footnote{In what follows, for the sake of simplicity we do not distinguish between $\bar\mu^2$ and $\mu^2$.}
\begin{align}\label{1loop-IR}
\F^{(1)}_\epsilon =M^{(1)}=  (x_{13}^2/\bar\mu^2)^{-\epsilon}\lr{\frac1{2\epsilon} +\frac12+ O(\epsilon^2)}\,.
\end{align}
Here the pole in $\epsilon$ comes from the integration over small
momenta $k\to 0$ and $(p-k)\to 0$ and, therefore, has an IR origin in the dual momentum representation.
Comparison of \p{1loop-IR} with the first relation in \p{F2} shows that
the singular term is reproduced correctly (upon  identifying $x_{12}^2, x_{34}^2\to \mu^2$ and
subtracting the pole),
while the regular (constant) term is different.

\subsection{Two loops}

Let us now extend the above analysis to two loops. According to \p{F2}, the two-loop correction to the correlation
function $\F^{(2)}$ has a stronger, $(\ln u)^2$ singularity for $u\to 0$. The reason for this is that, in the short-distance limit $x_1\to x_2$ and $x_3\to x_4$, the integrals on the right-hand side of \p{F12} develop overlapping singularities when the integration points $x_5$ and $x_6$ independently
approach the two external points, e.g. $x_5\to x_1$ and $x_6\to x_3$.  At the same time, the leading $(\ln u)^2$ singularity is supposed to cancel in the particular
combination of one- and two-loop corrections \p{spec}, which defines the $O(a^2)$ correction to the
logarithm of the correlation function on the left-hand side of \p{SD}.

To understand the reason for this, we replace
$\F^{(1)}$ and $\F^{(2)}$ on the left-hand side of \p{spec} by their explicit expressions \p{F12} and, then,
simplify the resulting expression by applying the same limiting procedure as in \p{F-F}. Namely, we take
the limit $x_1\to x_2$ and $x_3\to x_4$ inside the $g-$ and $h-$integrals and modify the integration measure
as in \p{scheme}. In this manner, we arrive at
\begin{align}\label{spec1}
\F_\epsilon^{(2)} - 3\, (\F_\epsilon^{(1)})^2 =
\lr{\frac{\mu^{2\epsilon}}{4\pi^2}}^2\int d^{4-2\epsilon} x_5d^{4-2\epsilon} x_6
\frac{2 x_{13}^6 (x_{15}^2x_{36}^2+x_{16}^2x_{35}^2-x_{13}^2x_{56}^2)}{(x_{15}^4x_{16}^4)x_{56}^2(x_{35}^4x_{36}^4)}\,,
\end{align}
where the expression on the right-hand side is manifestly symmetric with respect to the integration points, $x_5$ and $x_6$, and it
takes into account the contribution from the sum of $g^2-$ and $h-$integrals. It is clear from \p{spec1} that the
integral diverges logarithmically when $x_5$ and $x_6$ approach the external points $x_1$ and $x_3$. The
simplest way to evaluate \p{spec1} is by going to the dual momenta $k_1=x_{15}$, $k_2=x_{16}$ and $p=x_{13}$, so that
\begin{align}\label{2loop-IR}
\F_\epsilon^{(2)} - 3\, (\F_\epsilon^{(1)})^2 = 4M^{(2)} - 2 (M^{(1)})^2\,.
\end{align}
Here the one-loop integral $M^{(1)}$ was introduced in \p{I1} and $M^{(2)}$ stands for
the standard  two-loop scalar propagator-type integral \cite{MINCER}
\begin{align}\notag
M^{(2)} &= \lr{\frac{\mu^{2\epsilon}}{4\pi^2}}^2\int \frac{d^{4-2\epsilon} k_1 d^{4-2\epsilon} k_2}{k_1^4 k_2^2 (k_1-k_2)^2 (p-k_1)^2(p-k_2)^4}
\\[2mm]
\label{I2}
&
= (x_{13}^2/\mu^2)^{-2\epsilon}\lr{\frac1{8\epsilon^2} + \frac{3}{16\epsilon}-\frac1{16} + O(\epsilon)}\,.
\end{align}
Substituting \p{1loop-IR} and \p{I2} into \p{2loop-IR}, we find that the double pole cancels leading to
\begin{align}\label{2loop-eps}
\F_\epsilon^{(2)} - 3\, (\F_\epsilon^{(1)})^2 = (x_{13}^2/\mu^2)^{-2\epsilon}\lr{- \frac{1}{4\epsilon}-\frac3{4} + O(\epsilon)}\,.
\end{align}
We observe that the expansion of this expression around $\epsilon=0$ produces the logarithmic term
$-(1/2)\ln(\mu^2/x_{13}^2)$, which matches the $-(1/4)\ln u$ term on the right-hand side of \p{spec}
after the identification $x_{12}^2,x_{34}^2\to \mu^2$. As in the one-loop case, the finite terms in the two
expressions are different.  According to \p{ex1}, the coefficient in front of $\ln u$
is related to the two-loop Konishi anomalous dimension, $\gamma_\mathcal{K}^{(2)}/12$. Similarly to \p{gamma1},
this allows us to write
\begin{align}\label{K2}
\gamma_\mathcal{K}^{(2)} = 6\frac{d}{d\ln \mu^2} \big[{\F_\epsilon^{(2)} - 3\, (\F_\epsilon^{(1)})^2} \big]= -3\,,
\end{align}
in agreement with \p{Main}.

What is the reason why the double pole cancels in \p{spec1}? This happens because
the numerator in  \p{spec1} has the following characteristic feature: it vanishes  for $x_5\to x_1$ and $x_5\to x_3$ with $x_6$ in general position. As a consequence, the most singular contribution coming from the two regions, $x_5\to x_1, x_6\to x_3$ and $x_5\to x_3, x_6\to x_1$, is
suppressed and the integral in \p{spec1} develops a weaker singularity.
It only arises when the two integration points approach one of the external points simultaneously, $x_5, x_6\to x_1$
and $x_5, x_6\to x_3$. We can make use of this fact to further simplify the calculation of the divergent part of the
integral \p{spec1}.

Like in the one-loop case \p{delta}, we can single out the divergent contribution to  \p{spec1} by introducing a dimensionful parameter $\delta^2\ll x_{13}^2$ and restricting the integration in \p{spec1} to a ball of radius $\delta$ centred at the points $x_1$ or $x_3$. Due to the symmetry of the integral \p{spec1} under the exchange of $x_1$ and $x_3$, the two regions produce the same contribution leading to
\begin{align}\label{ball2} \notag
\F_\epsilon^{(2)} - 3\, (\F_\epsilon^{(1)})^2 &\sim 4
\lr{\frac{\mu^{2\epsilon}}{4\pi^2}}^2\int_{\Omega_\delta} d^{4-2\epsilon} x_5\,d^{4-2\epsilon} x_6
\frac{x_{15}^2 +x_{16}^2-x_{56}^2}{(x_{15}^4x_{16}^4)x_{56}^2}
\\
&=\frac12\lr{\frac{\mu^{2\epsilon}}{\pi^2}}^2\int_{\Omega_\delta} d^{4-2\epsilon} x_5\,d^{4-2\epsilon} x_6
\frac{  (x_{15}\cdot x_{16} )}{(x_{15}^4x_{16}^4)x_{56}^2}
\,.
\end{align}
Here the integration is performed over the region $\Omega_\delta$ defined as $x_{51}^2, x_{61}^2 <\delta^2$. Notice
that in this region we can safely replace $x_{35}^2$ and $x_{36}^2$ inside the integral with $x_{13}^2$.
To simplify \p{ball2}, we introduce the radial coordinates $r_5^2= x_{51}^2$, $r_6^2=x_{61}^2$ and the angle $\phi$ between the two vectors,  $(x_{15}\cdot x_{16} )=r_5 r_6 \cos\phi$. Integration over the angle yields~\cite{Chetyrkin:1980pr}
\begin{align}\label{pole2}
\text{r.h.s. of \p{ball2}} = \mu^{4\epsilon}\int_0^\delta \frac{d r_5 dr_6}{(r_5 r_6)^{2\epsilon}}\frac{r_<}{r_>^3} =
(\delta^2/\mu^2)^{-2\epsilon}\lr{- \frac{1}{4\epsilon}+ O(\epsilon^0)}\,,
\end{align}
where the notation was introduced for $r_<={\rm min}(r_5,r_6)$ and $r_>={\rm max}(r_5,r_6)$.
As expected, the residue of the pole in \p{pole2} is the same as in \p{2loop-eps}
and, therefore, it leads to the same result for the two-loop Konishi anomalous dimension \p{K2}.

\subsection{Loop reduction}

Comparing the integrals in Eqs.~\p{spec1} and \p{ball2}, we notice that the latter contains a smaller number of propagators and, therefore, is much easier to analyze.
Still, both integrals are two-loop ones as they involve integration over two points.
As we show in this subsection, there is yet another simplification which allows us to effectively
eliminate the integration over one point and, therefore, reduce the number of loops by one.

Let us return to the relation \p{ball2} and perform the integration over $x_6$ with the point $x_5$ in a general
position inside the region $\Omega_\delta$. It is easy to see that the $x_6-$integral depends on the two scales
$x_{15}^2$ and $\delta^2$, which play the role of cut-offs at short and large distances, respectively.
Notice that the integral converges at large $x_6$ and, therefore, it would stay finite if we sent $\delta^2/x_{15}^2$
to infinity. This means that, as far as the leading divergence of \p{ball2} is concerned, when computing the $x_6-$integral we can neglect its $\delta^2-$dependence and extend the integration over $x_6$ to the whole $(4-2\ep)-$dimensional
(Euclidean) space
\begin{align}\notag
\frac1{\pi^2}\int_{\Omega_\delta}  d^{4-2\epsilon} x_6
\frac{  (x_{15}\cdot x_{16} )}{x_{16}^4x_{56}^2} & = \frac1{\pi^2} \int  d^{4-2\epsilon} x_6
\frac{  (x_{15}\cdot x_{16} )}{x_{16}^4x_{56}^2}\times \left[1+ O(x_{15}^2/\delta^2)\right]
\\[2mm]\label{C}
&
= C_\epsilon (x_{15}^2)^{-\epsilon}\times \left[1+ O(x_{15}^2/\delta^2)\right].
\end{align}
Here the second relation follows from dimensional analysis of the integral and $C_\epsilon$ is some
constant regular at $\epsilon=0$. Substituting this result on the right-hand side of  \p{ball2} we
arrive at the integral
\begin{align}\label{ball3}
 \frac{C_\epsilon \mu^{4\epsilon}}{2\pi^2} \int_{\Omega_\delta} \frac{d^{4-2\epsilon} x_5}{(x_{15}^2)^{2+\epsilon}}\, \left[1+ O(x_{15}^2/\delta^2)\right]
\,.
\end{align}
It is easy to see that the $O(x_{15}^2/\delta^2)$ term  does not
produce a  pole in $\epsilon$ and, therefore,
can be safely neglected. Performing the $x_5-$integration, we finally obtain
\begin{align}\label{pole3}
\F_\epsilon^{(2)} - 3\, (\F_\epsilon^{(1)})^2 & \sim (\delta^2/\mu^2)^{-2\epsilon} \lr{-\frac{C_\epsilon}{4\epsilon}+ O(\epsilon^0)}\,.
\end{align}
Thus, the residue at the pole and, hence, the {\it two-loop} Konishi anomalous dimension,  is determined by the constant $C_\epsilon$ which is given in its turn by the {\it one-loop}
integral \p{C}
\begin{align}\label{C-eps}
C_\epsilon = \frac{(x_{15}^2)^{\epsilon}}{\pi^2} \int d^{4-2\epsilon} x_6
\frac{  (x_{15}\cdot x_{16} )}{x_{16}^4x_{56}^2}
=1 + O(\epsilon)\,.
\end{align}
Substituting this relation into \p{pole3}, we arrive at \p{pole2}.

\begin{figure}[t]
\psfrag{a}[cc][cc]{$\scriptstyle \alpha$}
\psfrag{b}[cc][cc]{$\scriptstyle \beta$}
\psfrag{c}[cc][cc]{$\scriptstyle \alpha+\beta-D/2$}
\psfrag{G}[cc][cc]{$ \times \ G(\alpha,\beta)$}
\centerline{\includegraphics[width=80mm]{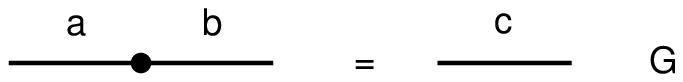}}
\caption{Diagrammatic representation of
the chain relation. Solid line with index $\alpha$ stands for $1/(x^2)^\alpha$ and black dot denotes the integration point.} \label{fig-chain}
\end{figure}

The key point in the above argument is that the final result in \p{C-eps} is given by an  integral which is {\it finite} for $\epsilon\to 0$.
Indeed, the potential singularities of this integral could come from integration in the vicinity of $x_1$ (at infinity it is obviously finite by power counting). However, the numerator of the integrand vanishes for $x_6 \to x_1$, so the integral is convergent. Notice that the relation \p{C-eps} can be rewritten in a form resembling  the first line in \p{ball2}:
\begin{align}\label{dimreg}
C_\epsilon = \frac{(x_{15}^2)^{\epsilon}}{2\pi^2} \left[\int \frac{d^{4-2\epsilon} x_6\,x^2_{15}}{ x^4_{16} x^2_{56}} + \int \frac{d^{4-2\epsilon} x_6}{x^2_{16} x^2_{56}} -  \int \frac{d^{4-2\epsilon} x_6}{x^4_{16}}\right]
  \,.
\end{align}
In this form, each integral in  the square brackets develops a pole $1/\epsilon$ but their sum is finite for $\epsilon\to 0$. The first integral diverges when $x_6 \to x_1$, the second one when $x_6 \to \infty$. Both of them can be easily evaluated
with the help of  the  ``chain relation'' \cite{Chetyrkin:1980pr} shown diagrammatically in Fig.~\ref{fig-chain}:
 \begin{align}\label{G-fun} \notag
& \frac1{\pi^{D/2}}\int   \frac{d^D x_0}{(x_{10}^2)^\alpha (x_{02}^2)^\beta} =  \frac{G(\alpha,\beta) }{(x_{12}^2)^{\alpha+\beta-D/2} }
\,,\
\\[2mm]
&
G(\alpha,\beta) = \frac{\Gamma(\alpha+\beta-D/2)\Gamma(D/2-\alpha)\Gamma(D/2-\beta)}{\Gamma(\alpha)\Gamma(\beta)\Gamma(D-\alpha-\beta)}\,.
\end{align}
It is easy to check that the pole $1/\epsilon$ cancels in the sum of two integrals. The third integral in \p{dimreg} is scaleless and, therefore, it vanishes in dimensional regularization.\footnote{More precisely, it develops two
poles $1/\ep$ after integration over $x_6\to x_1$ and $x_6\to\infty$, one of UV and the other of IR origin. Their residues have opposite signs, so they cancel in the sum. \label{foot}}

\subsection{The method}

We are now ready to formulate our method for computing the Konishi anomalous dimension at higher loops. It consists of four steps:

\medskip

{\bf Step 1:} Expand the logarithm of the correlation function \p{SD} in powers of the coupling constant and
obtain the $\ell-$loop correction to the left-hand side of \p{SD} in the form of  an $\ell-$folded integral over the
internal points
$x_5,\ldots,x_{4+\ell}$:
\begin{align}\label{I}
\ln  \lr{1+6 \sum_{\ell\ge 1} a^\ell \F^{(\ell)}(x_i)} =\sum_{\ell\ge 1} a^\ell
 \int d^4 x_5 \ldots d^4 x_{4+\ell}\,
\mathcal{I}_\ell(x_1,\ldots,x_4|x_5,\ldots,x_{4+\ell})\,,
\end{align}
with the integrand $\mathcal{I}_\ell$ symmetric under the $S_4\times S_\ell$ permutations
of the four external points, $x_1,\ldots,x_4$ and the $\ell$
integration points $x_5,\ldots,x_{4+\ell}$.
In the short-distance limit, for $x_1\to x_2$ and $x_3\to x_4$, the integral develops
a single logarithmic singularity.

\medskip

{\bf Step 2:}  Replace the integrand $\mathcal{I}_\ell$ by its limiting value at $x_1=x_2$ and $x_3=x_4$,
introduce the regularization \p{scheme} for the integration measure over the $\ell$ internal points and, then,
restrict the integration over $x_5,\ldots,x_{4+\ell}$ to a ball of radius $\delta$ centred at one of the
external points, say, $x_1$:
\begin{align}
2\times \sum_{\ell\ge 1} a^\ell (\mu^{2})^{\ell\epsilon} \int_{\Omega_\delta} d^{4-2\epsilon} x_5 \ldots d^{4-2\epsilon} x_{4+\ell}\,
\widehat{\mathcal{I}}_\ell (x_1|x_5\ldots,x_{4+\ell})\,.
\end{align}
Here  we inserted the factor of 2 to take into account the contribution from the integration around the point $x_3$,
and introduced the notation for
\begin{align}\label{I-hat}
\lim_{x_2\to x_1 \atop x_4\to x_3}\mathcal{I}_\ell(x_1,\ldots,x_4|x_5\ldots,x_{4+\ell})
= \widehat{\mathcal{I}}_\ell
(x_1|x_5\ldots,x_{4+\ell})+ O(\delta^2/ x_{13}^2)\,.
\end{align}

\medskip

{\bf Step 3:} Freeze one of the integration points, say, $x_5$ and perform the integration over the remaining $(\ell-1)$ points
by extending the integration region to the whole space:
\begin{align}\label{red}
2 \int d^{4-2\epsilon} x_6 \ldots d^{4-2\epsilon} x_{4+\ell}\,
\widehat{\mathcal{I}}_\ell (x_1|x_5\ldots,x_{4+\ell}) =\frac{C_{\ell-1}}{\pi^2}  (x_{15}^2)^{-2-(\ell-1)\epsilon}\,,
\end{align}
with the constant $C_{\ell-1}$  being regular at $\epsilon=0$. Going back to \p{I},
we perform the remaining $x_5-$integration and obtain:
\begin{align}\notag
\ln  \lr{1+6 \sum_{\ell\ge 1} a^\ell \F_\epsilon^{(\ell)}(x_i)} & \sim
\sum_{\ell\ge 1} a^\ell  C_{\ell-1} \frac{(\mu^{2})^{\ell\epsilon}}{\pi^2} \int_{\Omega_\delta} \frac{d^{4-2\epsilon} x_5}{(x_{15}^2)^{2+(\ell-1)\epsilon}}
\\
&=-\sum_{\ell\ge 1}  a^\ell \,{C_{\ell-1}}  \frac{1}{\ell\epsilon}  (\mu^{2}/\delta^2)^{\ell\epsilon} + O(\epsilon^0)\,.
\end{align}
The expansion of this relation in the powers of $\epsilon$ produces a $\ln\mu^2$ term which should match the $\ln u$ term on the right-hand side of  \p{SD} for $x_{12}^2,x_{34}^2\to \mu^2$.

\medskip

{\bf Step 4:} We compare the last relation with \p{SD},  identify $\ln u \to 2\ln(\mu^2/\delta^2)$ and obtain the Konishi anomalous dimension as
\begin{align}
 \gamma_\mathcal{K}(a) = \frac{d}{d\ln \mu^2}\ln  \lr{1+6 \sum_{\ell\ge 1} a^\ell \F_\epsilon^{(\ell)}(x_i)}
\end{align}
leading to
\begin{align}\label{K-C}
\gamma_\mathcal{K}(a) =  -\sum_{\ell\ge 1} a^\ell\,C_{\ell-1}\,.
\end{align}
This relation allows us to  express the Konishi anomalous dimension at $\ell$ loops in terms of the constants $C_{\ell-1}$ which are defined in their turn in terms of scalar integrals at $(\ell-1)$ loops, Eq.~\p{red}.

As we will show in this paper, going through these steps we will be able to determine the Konishi anomalous dimension
up to five loops in the planar sector, as well as at four loops in the non-planar sector (up to a rational prefactor, see below).

\section{Konishi anomalous dimension at three and four loops}

Let us apply the method described in the previous section to compute the Konishi anomalous dimension
at three loops and beyond.

\subsection{Preliminaries}

We start with the general expression for the logarithm of the correlation function in the short-distance limit on the left-hand side of \p{I},
\begin{align} \label{sum-I}
  \ln  \bigg(1& +6  \sum_{\ell\ge 1} a^\ell \F^{(\ell)}\bigg)  = \sum_{\ell\ge 1} a^\ell \, I^{(\ell)}\,,
\end{align}
where $I^{(\ell)}$ are given by the following expressions up to five loops:
\begin{align}\label{Is}
\notag
I^{(1)}& =  6\, \F^{(1)}\,,
\\ \notag
I^{(2)}& = 6 \big[\F^{(2)}-3(\F^{(1)})^{2}\big]\,,
\\ \notag
I^{(3)}& =6\big[\F^{(3)}-6\F^{(1)}\F^{(2)}+12(\F^{(1)})^{3}\big]\,,
\\ \notag
I^{(4)}& =6\big[\F^{(4)}-6\F^{(1)} \F^{(3)}-3(\F^{(2)})^{2}+36 \F^{(2)}(\F^{(1)})^{2}-54(\F^{(1)})^{4}\big]\,,
\\ \notag
I^{(5)}& =6\big[ \F^{(5)}-6 \F^{(1)}\F^{(4)}-6 \F^{(3)}\F^{(2)}+36 \F^{(3)}(\F^{(1)})^{2}
\\
& \hspace*{48mm}
+36 \F^{(1)}(\F^{(2)})^{2} -216\F^{(2)}(\F^{(1)})^{3}+{\ft {1296}{5}}(\F^{(1)})^{5} \big] \,.
\end{align}
We recall that the function $\F^{(\ell)}$ was defined in \p{F-hat} as the short-distance limit of the correlation function at $\ell$ loops, Eq.~\p{integg}.

To find $\F^{(\ell)}$, we substitute \p{integg} into \p{F-hat} and regularize the integration
measure according to \p{scheme}:
\begin{align} \label{F-hat-P}
  \F_\epsilon^{(\ell)}(x_i)={x_{13}^4\, \mu^{2\ell\epsilon} \over \ell!\,(-4\pi^2)^\ell} \int \frac{d^{4-2\epsilon}x_5 \dots
  d^{4-2\epsilon}x_{4+\ell}}{(x_{15}^4 x_{35}^4)\ldots (x_{1,4+\ell}^4 x_{3,4+\ell}^4)} \,  { \widehat{P}^{(\ell)}(x_1,x_3; x_5,\ldots
   x_{4+\ell}) \over \prod_{5\leq i<j \leq 4+\ell}   x_{ij}^2}\,.
\end{align}
Here $\widehat{P}^{(\ell)}$ coincides with the polynomial ${P}^{(\ell)}$, Eq.~\p{eq:10}, evaluated at $x_2=x_1$ and
$x_4=x_3$
\begin{align}\label{P-hat}
\widehat{P}^{(\ell)}=\lim_{x_2\to x_1\atop x_4\to x_3} {P}^{(\ell)}(x_1,\ldots,x_{4+\ell})\,.
\end{align}
For instance, for $\ell=1$ and $\ell=2$ we find from \p{P2}
\begin{align}
\widehat{P}^{(1)} =1\,,\qqqquad \widehat{P}^{(2)} = 2x_{13}^4 x_{56}^2 + 4 x_{13}^2 x_{15}^2 x_{36}^2 + 4 x_{13}^2 x_{16}^2 x_{35}^2 \,.
\end{align}
Notice that the polynomial ${P}^{(\ell)}$ is symmetric with respect to all $(4+\ell)$ points
while for the  polynomial $\widehat{P}^{(\ell)}$ this symmetry reduces to $S_2\times S_\ell$ permutations of two external points $x_1,x_3$ and of the $\ell$ integration points
$x_5,\ldots,x_{4+\ell}$.

Substituting the definition \p{F-hat-P} into \p{Is}, we can represent the right-hand side of \p{sum-I} in the same form as in \p{I}, in terms of the $\ell-$fold integrals
\begin{align}\label{I-ell}
 I^{(\ell)}  =  \mu^{2\ell\epsilon} \int d^{4-2\epsilon} x_5 \ldots d^{4-2\epsilon} x_{4+\ell}\
\mathcal{I}_\ell \,, 
\end{align}
and express the integrands $\mathcal{I}_\ell$ in terms of the polynomials $\widehat{P}^{(\ell')}$
with $1\le \ell'\le \ell$. For $\ell=1$ and $\ell=2$ we have
\begin{align} \notag
 {\mathcal{I}}_1 &= -\frac{3}{2\pi^2} \frac{x_{13}^4}{x_{15}^4 x_{35}^4}
\\ \label{P2-hat}
 {\mathcal{I}}_2 &=  \frac{3}{(4\pi^2)^2} \frac{x_{13}^4}{x_{15}^4 x_{35}^4x_{16}^4 x_{36}^4}\left(\frac{\widehat{P}_{5,6}}{x_{56}^2}-6 x_{13}^4  \right)
 =\frac{3}{4\pi^4}\frac{x_{13}^6(x_{15}^2x_{36}^2+x_{16}^2x_{35}^2-x_{13}^2x_{56}^2)}{x_{15}^4 x_{35}^4x_{16}^4 x_{36}^4 x_{56}^2}\,,
\end{align}
where $\widehat{P}_{5,6} \equiv \widehat{P}^{(2)}(x_5,x_6)$. It is easy to see that the numerator of ${\mathcal{I}}_2$ vanishes for $x_5\to x_1$ and $x_5\to x_3$,  as needed to achieve a lower degree of  divergence.

As was already explained, the integral in \p{I-ell} develops a {\it simple} pole $O(1/\epsilon)$ from
integration over all points $x_5,\ldots,x_{4+\ell}$ in the vicinity of the two external points
$x_1$ and $x_3$. For $x_i\to x_1$ we can safely replace $x_{3i}^2\to x_{13}^2$ inside
$\mathcal{I}_\ell$ without affecting the residue at the pole. In this way, we construct the function $\widehat{\mathcal{I}}_\ell$ defined in \p{I-hat}. At one and two loops we have
\begin{align}
\widehat{\mathcal{I}}_1= -\frac{3}{2\pi^2}\frac{1}{x_{15}^4}\,,\qqqquad
\widehat{\mathcal{I}}_2 = \frac{3}{4\pi^4}\frac{ (x_{15}^2 +x_{16}^2-x_{56}^2)}{x_{15}^4  x_{16}^4 x_{56}^2}\,.
\end{align}
By construction, the functions $\widehat{\mathcal{I}}_\ell$ do not depend on $x_3$. Substituting these relations in
\p{red}, we find with the help of \p{C-eps}
\begin{align}
C_0 =  -3\,,\qquad C_1= 3 C_\epsilon=3\,,
\end{align}
with $C_\epsilon$ defined in \p{C-eps}.

\subsection{Three loops}\label{sec:three-loops}

According to \p{integg} and \p{eq:10} the correlation function at three loops $F^{(3)}(x_i)$ is determined by the $S_7-$invariant polynomial $P^{(3)}(x_1,\ldots,x_7)$. As was shown
in Ref.~\cite{Eden:2012tu}, the form of this polynomial can be fixed from the requirement for the correlation function to have the correct asymptotic behaviour at short distances. The result is
\begin{align}
  P^{(3)}= \frac1{20}(x_{12}^2)^2( x_{34}^2 x_{45}^2 x_{56}^2 x_{67}^2 x_{73}^2)  \ +\ {S_7\ \mathrm{permutations}} ,
 \end{align}
where the sum runs over the $S_7$ permutations of the indices $1,\ldots,7$. The explicit
expression for $P^{(3)}$ contains 5040 distinct terms. However,
as was explained above, for our purposes we only need its limit \p{P-hat} for $x_2=x_1$ and $x_4=x_3$. This brings the number of terms down to 27:
\begin{align}\notag
 \widehat P^{(3)} & = 8 x_{13}^2 x_{16}^2 x_{17}^4 x_{35}^4 x_{36}^2
 + 4 x_{13}^4 x_{16}^2 x_{17}^2 x_{35}^2 x_{37}^2 x_{56}^2 + 4 x_{13}^6 x_{17}^2 x_{36}^2 x_{56}^2 x_{57}^2
 \\[2mm]
 &  + 2 x_{13}^4 x_{17}^4 x_{35}^2 x_{36}^2 x_{56}^2 + 2 x_{13}^4 x_{15}^2 x_{16}^2 x_{37}^4 x_{56}^2 + 2 x_{13}^6 x_{17}^2 x_{37}^2 x_{56}^4
 + \text{$S_3$ permutations}\,,
\end{align}
where the $S_3$ permutations only act on the integration points $x_5, x_6,x_7$. The polynomial
$\widehat P^{(3)}$ defined in this way is a completely symmetric function of
$x_5,x_6,x_7$.

Then, we apply \p{F-hat-P} and \p{Is} to define the integrand in \p{I-ell} at three loops:
\begin{align} \label{P3-hat}
 {\mathcal{I}}_3&= -\frac{1}{(4\pi^2)^3}\frac{x_{13}^4}{\prod_{i=5,6,7} x_{1i}^4 x_{3i}^4}
\left[\frac{\widehat{P}_{5,6,7}}{x_{56}^2x_{67}^2x_{75}^2}-6x_{13}^4
\lr{\frac{\widehat{P}_{5,6}}{x_{56}^2}+\frac{\widehat{P}_{6,7}}{x_{67}^2}+\frac{\widehat{P}_{5,7}}{x_{57}^2}} +72 x_{13}^8 \right],
\end{align}
where $\widehat{P}_{i,j}$ was defined in \p{P2-hat} and the notation was introduced for $\widehat{P}_{5,6,7} \equiv \widehat{P}^{(3)}(x_5,x_6,x_7)$.
Replacing the $\widehat{P}-$polynomials by their explicit expressions we get
\begin{align}\notag
 {\mathcal{I}}_3 =-\frac{1}{(4\pi^2)^3}\frac{2 x_{13}^6}{x_{15}^4 x_{16}^4 x_{17}^4 x_{35}^4 x_{36}^4 x_{37}^4
   x_{56}^2 x_{57}^2 x_{67}^2} \bigg[    3 x^6_{13}x^2_{56} x^2_{57} x^2_{67}  +x^4_{13} x^2_{17} x^2_{56} (x^2_{37} x^2_{56}-10 x^2_{36}
   x^2_{57})
\\
  +  x^2_{13}x^2_{56}\left(x^2_{35} x^2_{36} x^4_{17}+2 x^2_{16} x^2_{35} x^2_{37}
   x^2_{17}+x^2_{15} x^2_{16} x^4_{37}\right)  +4 x^2_{16} x^4_{17} x^4_{35}  x^2_{36}
   + \text{$S_3$ permutations}\bigg].
\end{align}
We verify that the numerator ${\mathcal{I}}_3$ vanishes when one of the integration points $x_5,x_6,x_7$
approaches the external points $x_1$ or $x_3$. This ensures that the integral in \p{I-ell}
develops a single pole only. It originates from two different integration regions,
$x_5,x_6,x_7\to x_1$ and $x_5,x_6,x_7\to x_3$, which produce however the same
contribution in virtue of the symmetry of ${\mathcal{I}}_3$ under the exchange of $x_1$ and $x_3$.

We examine ${\mathcal{I}}_3$ for $x_5,x_6,x_7\to x_1$ and determine the corresponding function \p{I-hat} at three loops
\begin{align}\notag
\widehat {\mathcal{I}}_3=-\frac{1}{(4\pi^2)^3}\frac{2}{x_{15}^4 x_{16}^4 x_{17}^4 x_{56}^2 x_{57}^2 x_{67}^2}\bigg[4 x_{15}^4x_{16}^2 +x_{15}^4 x_{67}^2 +x_{15}^2x_{67}^4+ 2 x_{15}^2x_{16}^2 x_{67}^2 +x_{15}^2 x_{16}^2 x_{56}^2 &
\\ \label{I3-hat}
-10x_{15}^2x_{56}^2x_{67}^2+3x_{56}^2x_{57}^2x_{67}^2+ \text{$S_3$ permutations }& \bigg] .
\end{align}
As before, $\widehat {\mathcal{I}}_3$ is a symmetric function of $x_5,x_6,x_7$ and it does not depend on $x_3$.
At the next step, we substitute the function $\widehat {\mathcal{I}}_3$ into the left-hand side of \p{red} and integrate it over $x_6,x_7$ with  $x_5$ fixed
\begin{align}\label{C2-int}
  \int d^{4-2\epsilon} x_6 d^{4-2\epsilon} x_{7}\,
\widehat{\mathcal{I}}_3(x_1|x_5,x_6,x_7)  =\frac{C_{2}}{2\pi^2}  (x_{15}^2)^{-2-2\epsilon}\,.
\end{align}

In close analogy with the situation at two loops (see Eq.~\p{dimreg} and the discussion around it),  the integral on the left-hand side of \p{C2-int} is finite for $\epsilon\to 0$, due to the special
properties of the expression in the square brackets in \p{I3-hat}. We would like to emphasize that this only holds for the sum of all terms
in the square brackets, but not for each individual term. In other words, if we
split the integral on the left-hand side of \p{C2-int} into a sum of integrals corresponding to each terms in \p{I3-hat},
then each integral develops a pole $1/\epsilon$.
The poles cancel in the sum of all integrals leading to a finite expression for the coefficient
$C_2$ in \p{C2-int}.~\footnote{To be more precise, the distribution $(x_{15}^2)^{-2-2\ep}$ is singular for $\ep \to 0$. So, the left-hand side of \p{C2-int} has a pole $1/\epsilon$ whose residue is a contact term, see Appendix A.}
As an example, let us examine the integrals corresponding to the two terms in the second
line of \p{I3-hat}:
\begin{align}\notag
\int \frac{d^{4-2\epsilon} x_6 d^{4-2\epsilon} x_{7}}{x_{15}^4 x_{16}^4 x_{17}^4 x_{56}^2 x_{57}^2 x_{67}^2} & \big[-10x_{15}^2x_{56}^2x_{67}^2+3x_{56}^2x_{57}^2x_{67}^2 \big]
\\ \label{zero}
& \hspace*{-10mm}
=
-\frac{10}{x_{15}^2}\int \frac{d^{4-2\epsilon} x_6}{x_{16}^4} \int\frac{d^{4-2\epsilon} x_7}{x_{17}^4   x_{57}^2  } + \frac{3}{x_{15}^4}\int\frac{d^{4-2\epsilon} x_6}{x_{16}^4} \int\frac{d^{4-2\epsilon} x_7}{x_{17}^4} =0\,,
\end{align}
where in the second relation we took into account that the integral over $x_6$ is scaleless and, therefore, it vanishes in dimensional regularization (see footnote \ref{foot}).

\begin{figure}[ht]
\psfrag{2}[cc][cc]{$\scriptstyle 2$}
\psfrag{x1}[cc][cc]{$x_1$}\psfrag{x7}[cc][cc]{$x_5$}
\centerline{\includegraphics[width=\linewidth]{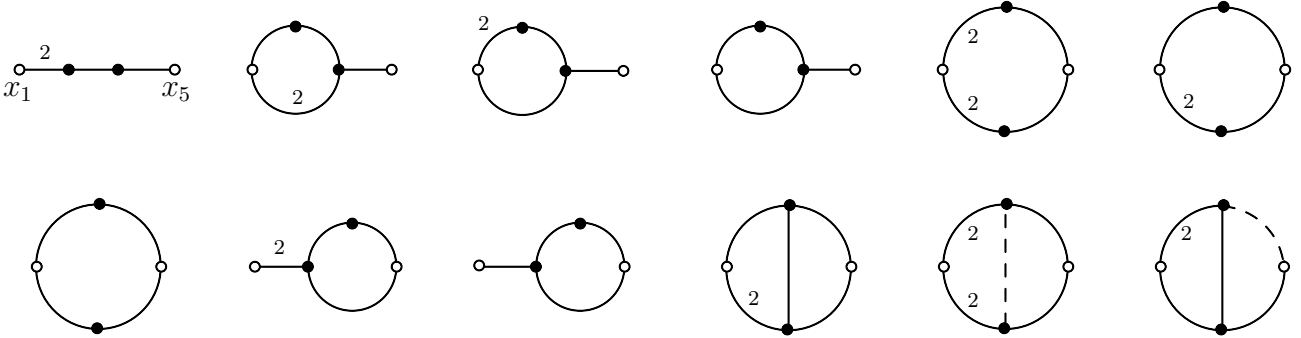}}
\caption{Diagrammatic representation of the
integrals in  \p{C2-fin}.
Solid lines without labels depict propagators $1/x_{ij}^2$, the label 2 refers to the square of the propagator. Dashed lines represent numerator factors $x_{ij}^2$, black and white dots represent  integration and external points, respectively. } \label{fig-3loop}
\end{figure}

Examining the contribution to \p{C2-int} from the remaining terms on the right-hand side
of \p{I3-hat}, we find that most of the integrals vanish in the same manner as in \p{zero}.
The remaining non-zero contribution  takes the following form:
\begin{align}\notag
C_2=-\frac{1}{8\pi^4}\int  d^{4-2\epsilon} x_6 d^{4-2\epsilon} x_{7}\bigg[   \frac{2}{x_{17}^4 x_{56}^2x_{67}^2}
+  \frac2{x_{16}^4x_{17}^2 x_{56}^2x_{67}^2}+\frac2{x_{16}^2x_{17}^4x_{56}^2x_{67}^2}+\frac2{x_{16}^2x_{17}^2x_{56}^2x_{67}^2}&
\\\notag
+  \frac1{x_{16}^4x_{17}^4 x_{56}^2x_{57}^2}+\frac2{x_{16}^4x_{17}^2 x_{56}^2x_{57}^2}
+\frac1{x_{16}^2x_{17}^2 x_{56}^2x_{57}^2}+\frac4{x_{16}^4x_{56}^2x_{57}^2x_{67}^2}&
\\ \label{C2-fin}
+ \frac4{x_{16}^2 x_{56}^2x_{57}^2x_{67}^2}+\frac4{x_{16}^4x_{17}^2x_{67}^2x_{57}^2x_{56}^2}+\frac{x_{67}^2}{x_{16}^4 x_{17}^4x_{56}^2x_{57}^2}+\frac{2x_{57}^2}{x_{16}^2x_{17}^4x_{56}^2x_{67}^2} & \bigg]\,.
\end{align}
Here we set $x_{15}^2=1$ for simplicity since the dependence of the integral on $x_{15}^2$ is uniquely fixed by dimension analysis. It is convenient to represent the 12
integrals in this relation  in the form of Feynman diagrams as shown in Fig.~\ref{fig-3loop}.

We observe that the first 9 integrals in \p{C2-fin} can be easily evaluated
with the help of  the  ``chain relation''  \p{G-fun} shown in Fig.~\ref{fig-chain}.
 Applying \p{G-fun} consecutively,  we can express the first 9 integrals in terms of the $G-$function and obtain the following relation (with $D=4-2\epsilon$)
 \begin{align}\label{C2-G} \notag
C_2=-\frac18\bigg[&  2 G(1,1) G(2,2-D/2) +2 G(1,1)G(1,4-D/2) +2 G(2,1)G(1,4-D/2)
\\\notag
+&
2G(1,1)G(1,3-D/2) + (G(2,1))^2 +2 G(1,1) G(2,1) +(G(1,1))^2
\\
+&4 G(1,1) G(2,3-D/2)
+ 4 G(1,1)G(1,3-D/2) + I_{10} + I_{11} + I_{12}\bigg] .
\end{align}
Here $I_{10}$, $I_{11}$ and $I_{12}$ stand for the last three integrals on the right-hand side of \p{C2-fin}, corresponding to the last three diagrams in Fig.~\ref{fig-3loop}. The simplest way to compute them is to introduce the dual momenta $k_1=x_{16}$, $k_2=x_{17}$ and $p=x_{15}$ and rewrite the above integrals as {\it two-loop propagator-type} momentum integrals. The latter
can be easily computed using {\tt MINCER}  \cite{MINCER}, yielding
\begin{align}\notag
I_{10} &= 4 (G(1,1))^2\left(\frac12 - \frac52\epsilon   + \frac92 \epsilon^2 + O(\epsilon^3)\right)\,,
\\[2mm]  \notag
I_{11} &= (G(1,1))^2\left( - 2  + 4 \epsilon-2\epsilon^2+ O(\epsilon^3)\right)\,,
\\[3mm]
I_{12}& =2(G(1,1))^2\left(1  -2 \epsilon + 4 \epsilon^2+ O(\epsilon^3)\right)\,.
\end{align}
Substituting these relations in \p{C2-G} we verify that all poles in $\epsilon$ cancel
on the right-hand side of \p{C2-G}, leading to the Konishi anomalous
dimension at three loops, Eq.~\p{K-C},
\begin{align}
\gamma_\mathcal{K}^{(3)} =-C_2= \frac{21}4 \,,
\end{align}
in agreement with \p{Main}.

\subsection{Four loops}

A novel feature that we first encounter at four loops is that the correlation function $F^{(4)}$ receives non-planar corrections
\begin{align}
F^{(4)} = F^{(4)}_{\rm g=0} + \frac1{N_c^2} F^{(4)}_{\rm g=1}\,.
\end{align}
The two functions $F^{(4)}_{\rm g=0}$ and $F^{(4)}_{\rm g=1}$ have the same general form \p{integg} and \p{eq:10} and are defined by the polynomial
\begin{align}
 P^{(4)}(x_1,\ldots,x_8) = P^{(4)}_{\rm g=0} + \frac1{N_c^2} P^{(4)}_{\rm g=1}\,.
\end{align}
 By construction, the  polynomials $P^{(4)}_{\rm g=0}$ and $P^{(4)}_{\rm g=1}$ are invariant under $S_8-$permutations of the four external points
 $x_1,\ldots,x_4$ and the four integration points $x_5,\ldots,x_8$.

Similarly to three loops, the explicit form of the planar polynomial $P^{(4)}_{\rm g=0}$ can be found from the requirement for the correlation function in the planar sector to have the correct asymptotic behaviour in the light-cone
limit $x_{12}^2,x_{23}^2,x_{34}^2,x_{41}^2\to 0$. The result is \cite{Eden:2012tu}
\begin{align} \notag \label{P-4loop-p}
P^{(4)}_{\rm g=0}& =\ft1{24} x_{12}^2 x_{13}^2 x_{16}^2 x_{23}^2 x_{25}^2 x_{34}^2 x_{45}^2 x_{46}^2 x_{56}^2 x_{78}^6
\\[2mm] \notag
& +\ft18x_{12}^2 x_{13}^2 x_{16}^2
   x_{24}^2 x_{27}^2 x_{34}^2 x_{38}^2 x_{45}^2 x_{56}^4 x_{78}^4
\\[2mm]
 & -\ft1{16} x_{12}^2 x_{15}^2 x_{18}^2 x_{23}^2 x_{26}^2 x_{34}^2
   x_{37}^2 x_{45}^2 x_{48}^2 x_{56}^2 x_{67}^2 x_{78}^2+\text{$S_8$ permutations}\,.
\end{align}
In the non-planar sector, the same requirement turns out to be less restrictive but it allowed us to determine the
non-planar polynomial $P^{(4)}_{\rm g=1}$ up to four arbitrary constants which are expected to take rational values. In order to fix these constants, one would need
more detailed information about the properties of the correlation function. To save space, here we do not present the general expression for  $P^{(4)}_{\rm g=1}$, it can be found in Ref.~\cite{Eden:2012tu}.

In fact, for our purposes we only need the expressions for
the polynomials $P^{(4)}_{\rm g=0}$ and $P^{(4)}_{\rm g=1}$  in the short-distance limit $x_2\to x_1$ and $x_4\to x_3$, Eq.~\p{P-hat}.  For the planar polynomial we find from \p{P-4loop-p} for
$x_2=x_1$ and $x_4=x_3$
\begin{align}\notag
\widehat P^{(4)}_{\rm g=0}& = x_{13}^8 \big(x_{15}^2 x_{18}^2 x_{36}^2 x_{37}^2 x_{58}^4
   x_{67}^4  +x_{16}^2 x_{17}^2 x_{36}^2
   x_{37}^2 x_{58}^6 x_{67}^2
  +4 x_{17}^2
   x_{18}^2 x_{36}^2 x_{37}^2 x_{56}^2 x_{58}^4
   x_{67}^2
\\  \notag
&
   +4 x_{16}^2 x_{17}^2 x_{37}^2
   x_{38}^2 x_{56}^2 x_{58}^4 x_{67}^2
   +2
   x_{17}^2 x_{18}^2 x_{37}^2 x_{38}^2 x_{56}^4
   x_{58}^2 x_{67}^2  +4 x_{17}^2 x_{18}^2
   x_{36}^2 x_{38}^2 x_{56}^2 x_{57}^2 x_{58}^2
   x_{67}^2
\\   \notag
&
   +2 x_{16}^2 x_{18}^2 x_{35}^2 x_{37}^2 x_{57}^2 x_{58}^2 x_{67}^2 x_{68}^2 x_{13}^8\big)
   +2 x_{13}^6\big( x_{17}^2 x_{18}^4 x_{37}^4 x_{38}^2 x_{56}^6
   +2 x_{17}^4 x_{18}^2 x_{36}^2 x_{38}^4 x_{56}^4 x_{57}^2
\\   \notag
&
   +2 x_{16}^2 x_{18}^4 x_{37}^4 x_{38}^2 x_{56}^4 x_{57}^2
   +2 x_{17}^2 x_{18}^4 x_{36}^4 x_{37}^2 x_{56}^2 x_{57}^2 x_{58}^2
   +2 x_{15}^2 x_{18}^4 x_{36}^2 x_{37}^4 x_{56}^2 x_{58}^2 x_{67}^2 \big)
\\   \notag
&
   +2x_{13}^4\big( x_{17}^2 x_{18}^6 x_{35}^2 x_{36}^2 x_{37}^4 x_{56}^4
   + x_{15}^2 x_{16}^2 x_{18}^4 x_{37}^6 x_{38}^2 x_{56}^4
   +2 x_{16}^2 x_{17}^2 x_{18}^4 x_{35}^2x_{37}^4 x_{38}^2 x_{56}^4
\\   \notag
&
   +2 x_{15}^2 x_{16}^2 x_{17}^4 x_{36}^2 x_{38}^6 x_{56}^2x_{57}^2
   +2 x_{15}^2 x_{17}^4 x_{18}^2 x_{36}^4 x_{38}^4 x_{56}^2 x_{57}^2
   +2 x_{17}^2 x_{18}^6 x_{35}^2 x_{36}^4 x_{37}^2 x_{56}^2 x_{57}^2
\\   \notag
&
   +2 x_{17}^4 x_{18}^4 x_{35}^2 x_{36}^4 x_{38}^2 x_{56}^2 x_{57}^2
   -  x_{16}^2 x_{17}^2 x_{18}^4 x_{35}^4 x_{36}^2 x_{37}^2 x_{58}^2 x_{67}^2\big)
   +4 x_{13}^2\big(x_{15}^2 x_{16}^2 x_{18}^6 x_{35}^2 x_{36}^2 x_{37}^6 x_{56}^2
\\   \notag
&
   +2 x_{16}^2 x_{17}^2
   x_{18}^6 x_{35}^4 x_{36}^2 x_{37}^4 x_{56}^2
   +2 x_{15}^2 x_{16}^4 x_{18}^4 x_{35}^2
   x_{37}^6 x_{38}^2 x_{56}^2 +2 x_{16}^4
   x_{17}^2 x_{18}^4 x_{35}^4 x_{37}^4 x_{38}^2
   x_{56}^2
\\    \label{P4-g=0}
&
   - x_{15}^2 x_{16}^2 x_{17}^2
   x_{18}^4 x_{35}^2 x_{36}^2 x_{37}^4 x_{38}^2
   x_{56}^2  \big) + \text{$S_4$ permutation}\,.
\end{align}
Here the $S_4$ permutations  are
needed to restore the symmetry of  $\widehat P^{(4)}_{\rm g=0}$ under the exchange of
the integration points $x_5,\ldots,x_8$. The polynomial $\widehat P^{(4)}_{\rm g=0}$ is also invariant under the
exchange of the external points $x_1$ and $x_3$, so that it has an $S_2\times S_4$
permutation symmetry. Notice that the relatively long expression for $\widehat P^{(4)}_{\rm g=0}$, as compared with \p{P-4loop-p}, is an artifact of decomposing the $S_8$ permutations
into $S_4\times S_4$ ones.

For the non-planar polynomial the situation is just the opposite. The expression for $P^{(4)}_{\rm g=1}$ is very lengthy whereas in the short-distance limit it takes a remarkably simple form. Namely, the four different polynomials that accompany the four arbitrary constants in the expression
for $P^{(4)}_{\rm g=1}$ become proportional to each other at $x_2=x_1$ and $x_4=x_3$, leading to~\cite{Eden:2012tu}
\begin{align}\label{P-g=1}
\widehat P^{(4)}_{\rm g=1} = \lim_{x_1\to x_2\atop x_3\to x_4} P_{\rm g=1}^{(4)} & = \frac16 c_{\rm g=1}^{(4)}
\, x_{13}^4
\lr{x_{56}^2x_{78}^2+ x_{57}^2x_{68}^2+x_{58}^2x_{67}^2}
\prod_{i=5,6,7,8} x_{1i}^2 x_{3i}^2\,,
\end{align}
with $c_{\rm g=1}^{(4)}$ given by a linear combination of the four rational constants mentioned above.

Then, we substitute the polynomial $\widehat P^{(4)}$ into \p{F-hat-P} and use \p{Is} to
obtain the expression for $I^{(4)}$. Going  to the corresponding integrand $\mathcal{I}_4$, Eq.~\p{I-ell}, we find
\begin{align}
\mathcal{I}_4 =  \mathcal{I}_{4,\rm g=0} + \frac1{N_c^2} \mathcal{I}_{4,\rm g=1}\,.
\end{align}
Here, the expression for the non-planar correction reads
\begin{align}\notag
 \mathcal{I}_{4,\rm g=1} &= {6 \over 4!\,(4\pi^2)^4} \frac{x_{13}^4}{(x_{15}^4 x_{35}^4)\ldots (x_{18}^4 x_{38}^4)} \,  { \widehat{P}_{\rm g=1}^{(4)}(x_5,\ldots
   x_{8}) \over \prod_{5\leq i<j \leq 8}   x_{ij}^2}
\\\label{I4-np}
&=   {c_{\rm g=1}^{(4)}  \over 4!\,(4\pi^2)^4} \frac{x_{13}^8}{(x_{15}^2 x_{35}^2)\ldots (x_{18}^2 x_{38}^2)} \,  { \lr{x_{56}^2x_{78}^2+ x_{57}^2x_{68}^2+x_{58}^2x_{67}^2}\over
x_{56}^2x_{57}^2x_{58}^2 x_{67}^2 x_{68}^2x_{78}^2}\,,
\end{align}
where in the second relation we replaced $\widehat{P}_{\rm g=1}^{(4)}$ by its explicit
expression \p{P-g=1}. In the planar sector, the analogous expression for $\mathcal{I}_{4,\rm g=0}$ is much longer
since it involves the $\widehat{P}-$polynomials at lower loops:
\begin{align}\notag
 \mathcal{I}_{4,\rm g=0} = \frac{6}{4! (4\pi^2)^4}&\frac{x_{13}^4}{\prod_{i=5}^8 x_{1i}^4 x_{3i}^4}\bigg[  \frac1{4!}\frac{\widehat{P}_{5,6,7,8}}{x_{56}^2x_{57}^2x_{58}^2 x_{67}^2 x_{68}^2x_{78}^2} - x_{13}^4\frac{  \widehat{P}_{5,6,7}}{x_{56}^2 x_{57}^2 x_{67}^2}
 \\[2mm]
 & \label{I4-p}
  -\frac34 x_{13}^4 \frac{ \widehat{P}_{5,6}\widehat{P}_{7,8}}{x_{56}^2 x_{78}^2} +18(x_{13}^4)^2 \frac{ \widehat{P}_{5,6}}{x_{56}^2} -54 (x_{13}^4)^{3}\bigg]+ \text{$S_4$ permutations}
\end{align}
where $\widehat{P}_{5,6,7,8}\equiv \widehat{P}_{\rm g=0}^{(4)}$ is given by \p{P4-g=0} and the right-hand side is symmetrized with respect to all $S_4$ permutations
of the integration points $x_5,\ldots,x_8$. Replacing the $ \widehat{P}-$polynomials
by their explicit expressions, we obtain a lengthy result for $\mathcal{I}_{4,\rm g=0}$.

At the next step, we restrict all the integration points $x_5,\ldots,x_8$ to  the
vicinity of the external point $x_1$ and simplify the integrand $\mathcal{I}_4$ by
replacing $x_{i3}^2\to x_{13}^2$ (with $i=5,\ldots,8$). Denoting the resulting
function $ \widehat{\mathcal{I}}_4$, we find from  \p{I4-np} and \p{I4-p}
\begin{align}\label{npl}
 \widehat{\mathcal{I}}_{4,\rm g=1} &=   {c_{\rm g=1}^{(4)}  \over 4!\,(4\pi^2)^4}    {  {x_{56}^2x_{78}^2+ x_{57}^2x_{68}^2+x_{58}^2x_{67}^2}\over  x_{15}^2 x_{16}^2 x_{17}^2 x_{18}^2
x_{56}^2x_{57}^2x_{58}^2 x_{67}^2 x_{68}^2x_{78}^2}
\end{align}
in the non-planar sector, and
\begin{align} \notag
 \widehat{\mathcal{I}}_{4,\rm g=0}=& \frac{6}{4! (4\pi^2)^4}{1 \over  x_{15}^4 x_{16}^4 x_{17}^4 x_{18}^4
x_{56}^2x_{57}^2x_{58}^2 x_{67}^2 x_{68}^2x_{78}^2}
 \\ \notag
 \times &\big[
 2 x_{17}^2 x_{56}^4 x_{18}^6
 +4 x_{15}^2x_{16}^2
   x_{56}^2x_{18}^6
   +8 x_{16}^2x_{17}^2x_{56}^2
   x_{18}^6
   +4 x_{17}^2x_{56}^2x_{57}^2x_{18}^6+2
   x_{17}^2x_{56}^6 x_{18}^4
 \\ \notag &  +2 x_{15}^2x_{16}^2 x_{56}^4 x_{18}^4
   +4 x_{16}^2x_{17}^2x_{56}^4
   x_{18}^4
   +8 x_{15}^2x_{16}^4 x_{56}^2x_{18}^4
   +8 x_{16}^4 x_{17}^2x_{56}^2x_{18}^4
\\ \notag &   -4 x_{15}^2
   x_{16}^2x_{17}^2x_{56}^2x_{18}^4
   +4 x_{16}^2
   x_{56}^4 x_{57}^2x_{18}^4
   +4 x_{17}^4 x_{56}^2
   x_{57}^2x_{18}^4
   -44 x_{17}^2x_{56}^2x_{57}^2
   x_{58}^2x_{18}^4
\\  \notag&    -2 x_{16}^2x_{17}^2x_{58}^2
   x_{67}^2x_{18}^4
   +4 x_{15}^2x_{56}^2x_{58}^2
   x_{67}^2x_{18}^4
   -12 x_{56}^2x_{57}^2x_{58}^2
   x_{67}^2x_{18}^4
   +x_{15}^2x_{58}^4 x_{67}^4
   x_{18}
\\  \notag&     -12 x_{56}^2x_{57}^2x_{58}^2x_{67}^4
   x_{18}
   +4 x_{17}^4 x_{56}^4 x_{57}^2x_{18}^2
   +4
   x_{15}^2x_{17}^4 x_{56}^2x_{57}^2x_{18}^2
   +4
   x_{17}^2x_{56}^2x_{58}^4 x_{67}^2x_{18}^2
\\ \notag &     +2
   x_{17}^2x_{56}^4 x_{58}^2x_{67}^2x_{18}^2
   -20
   x_{17}^2x_{56}^2x_{57}^2x_{58}^2x_{67}^2
   x_{18}^2
   -46 x_{16}^2x_{57}^2x_{58}^2x_{67}^2
   x_{68}^2x_{18}^2
\\ \notag &     +96 x_{56}^2x_{57}^2x_{58}^2
   x_{67}^2x_{68}^2x_{18}^2
   +4 x_{15}^2x_{16}^2
   x_{17}^4 x_{56}^2x_{57}^2
   +x_{16}^2x_{17}^2
   x_{58}^6 x_{67}^2
   +4 x_{16}^2x_{17}^2x_{56}^2
   x_{58}^4 x_{67}^2
\\ &     -12 x_{16}^2x_{17}^2x_{56}^2
   x_{57}^2x_{58}^2x_{67}^2
   -21 x_{56}^2x_{57}^2
   x_{58}^2x_{67}^2x_{68}^2x_{78}^2\big]+ \text{$S_4$ permutation}
\label{I4hat-p}
\end{align}
in the planar sector. We verify that the functions $\widehat{\mathcal{I}}_{4,\rm g=0}$ and $\widehat{\mathcal{I}}_{4,\rm g=1}$ transform covariantly under conformal transformations and
do not depend on the external point $x_3$.

Then, to find the four-loop correction to the Konishi anomalous dimension \p{K-C}, or equivalently the coefficient $C_3$, we substitute $\widehat{\mathcal{I}}_4$ into \p{red} and obtain:
 \begin{align} \label{C3}
C_3= \frac{2\pi^2}{(x_{15}^2)^{2+3\epsilon}} \int d^{4-2\epsilon} x_6 d^{4-2\epsilon} x_7 d^{4-2\epsilon} x_{8}\,\lr{
\widehat{\mathcal{I}}_{4,\rm g=0} + \frac1{N_c^2} \widehat{\mathcal{I}}_{4,\rm g=1}} \,.
\end{align}
By  construction,  $C_3$ is dimensionless and is expected to be finite as $\epsilon\to 0$. We do not remove the regularization, however, since it is more advantageous to expand
the right-hand side of \p{C3} into a sum of basis three-loop (divergent) integrals and
evaluate each of them separately.

\begin{figure}[h]
\psfrag{=}[cc][cc]{$=$}
\psfrag{x1}[cc][cc]{$x_1$}\psfrag{x5}[cc][cc]{$x_5$}
\psfrag{x6}[cc][cc]{$x_6$}\psfrag{x7}[cc][cc]{$x_7$}\psfrag{x8}[cc][cc]{$x_8$}
\centerline{\includegraphics[width=0.5\linewidth]{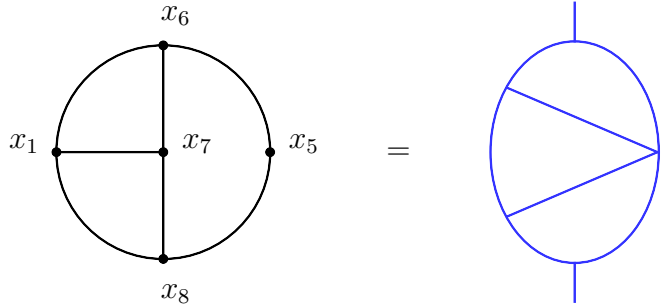}}
\caption{Diagrammatic representation of the integral \p{C3-np1} in the $x-$representation (left) and in the
dual momentum representation (right). {In what follows, all momentum integrals are shown in blue.} }
\label{fig-np}
\end{figure}

Let us start with the non-planar contribution to \p{C3}. Setting for simplicity  $x_{15}^2=1$, we
find from  \p{npl}, \p{C3}
\begin{align}\label{C3-np1}
C_{3,\rm g=1}=   {c_{\rm g=1}^{(4)}  \over 1024 \pi^6}   \int
    {  {d^{4-2\epsilon} x_6 d^{4-2\epsilon} x_7 d^{4-2\epsilon} x_{8}}\over x_{16}^2 x_{17}^2 x_{18}^2
 x_{56}^2x_{58}^2 x_{67}^2 x_{78}^2 }\,.
\end{align}
The integral on the right-hand side of this relation is shown diagrammatically in Fig.~\ref{fig-np}. It is finite at $\epsilon=0$ and, most importantly, it corresponds to a planar graph.
This fact allows us to introduce  dual momenta and represent the   integral in the
form of a planar dual (momentum) graph, as shown in  Fig.~\ref{fig-np}. 
The main advantage
of using the dual representation is that the resulting three-loop momentum integral
can be easily evaluated using the {\tt MINCER} package \cite{MINCER}, leading to
\begin{align}\label{C3-np}
C_{3,\rm g=1}=   {c_{\rm g=1}^{(4)}  \over 1024}\times \lr{20\zeta_5 +O(\epsilon)}\,.
\end{align}
We recall that the constant  $c_{\rm g=1}^{(4)}$ is expected to take rational values.

Let us now examine the three-loop integrals in \p{C3} generated by the integrand $\widehat{\mathcal{I}}_{4,\rm g=0}$, Eq.~\p{I4hat-p}. We shall denote the corresponding contribution
to the right-hand side of \p{C3} by $C_{3,\rm g=0}$.
The three-loop propagator integrals appearing in the calculation belong to the
following family of three-fold integrals, with various integer
indices $a_1,\ldots,a_{9}$
\begin{align}
 G(a_1,\ldots,a_9)=\int
\frac{d^{4-2\epsilon} x_6 d^{4-2\epsilon} x_7 d^{4-2\epsilon} x_8}{
 (x_{16}^2)^{a_1} (x_{17}^2)^{a_2}(x_{18}^2)^{a_3}
(x_{6}^2)^{a_4} (x_{7}^2)^{a_5}(x_{8}^2)^{a_6}  (x_{67}^2)^{a_7} (x_{68}^2)^{a_8} (x_{78}^2)^{a_9} }\,,
 \label{prop3l}
\end{align}
where  we put $x_5=0$ for simplicity. Notice that the  indices $a_i$ can take both positive
and negative values. In the latter case, the corresponding term
appears in the numerator. For arbitrary choices of the indices  $a_i$
some of the integrals \p{prop3l} may be non-planar and so cannot be rewritten as dual momentum integrals of propagator type. However, most importantly, in our case all the integrals appearing in \p{C3} are planar and thus of propagator type.

\begin{figure}[t]
\psfrag{L1}[cc][cc]{$L_{1}$}
\psfrag{P1}[cc][cc]{$P_1$}
\psfrag{P2}[cc][cc]{$P_2$}
\psfrag{P3}[cc][cc]{$P_3$}
\psfrag{P4}[cc][cc]{$P_4$}
\centerline{\includegraphics[width=0.95\linewidth]{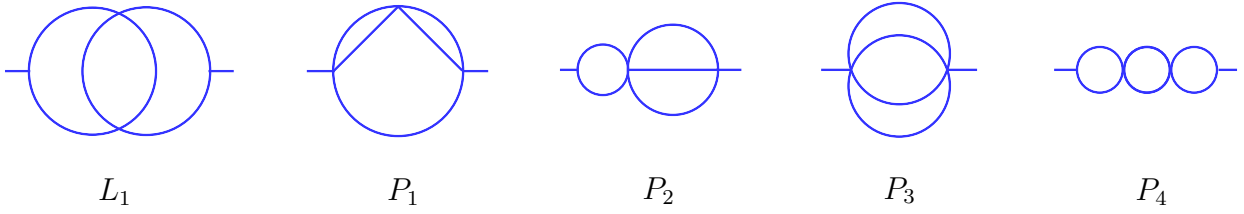}}
\caption{The dual momentum master integrals defining the four-loop Konishi anomalous dimension.}
\label{fig-basis-K}
\end{figure}

To evaluate each of them we apply the standard technique called integration by parts (IBP)
\cite{IBP} (see Chapter~5 of \cite{VSbooks} for a review of the method)
which provides the possibility of representing a given integral of this family
as a linear combination of so-called master integrals. We found that the calculation of $C_{3,\rm g=0}$ involves only 5 master integrals, all corresponding to planar graphs.
Therefore, introducing the dual momenta $k_i=x_i-x_{i+1}$, we can rewrite these integrals as the dual momentum-space integrals $L_1,P_1,\ldots,P_4$ shown in Fig.~\ref{fig-basis-K}.

The resulting
expression for  $C_{3,\rm g=0}$ is
\begin{align} \label{C3-IBP} \notag
C_{3,\rm g=0}& = L_1\left( \frac{9}{64 \epsilon ^2}-\frac{3}{4 \epsilon }+\frac{45}{16}-\frac{165 \epsilon }{16}+\frac{1647 \epsilon ^2}{64}-54
   \epsilon ^3+O\left(\epsilon ^4\right)\right)
\\ & \notag
+P_1\left( -\frac{9}{32 \epsilon ^3}+\frac{93}{32 \epsilon ^2}-\frac{981}{64 \epsilon }+\frac{2967}{64}-\frac{4035 \epsilon
   }{32}+\frac{2961 \epsilon ^2}{8}+O\left(\epsilon ^3\right)\right)
\\ & \notag
+P_2\left( \frac{9}{16 \epsilon ^3}-\frac{99}{32 \epsilon ^2}+\frac{195}{16 \epsilon }-\frac{1815}{32}+\frac{1749 \epsilon
   }{8}-\frac{9585 \epsilon ^2}{16}+O\left(\epsilon ^3\right)\right)
\\ & \notag
+P_3\left(\frac{27}{4 \epsilon ^3}-\frac{1035}{16 \epsilon ^2}+\frac{3267}{32 \epsilon }+\frac{8415}{16}-\frac{62247 \epsilon
   }{32}+O\left(\epsilon ^2\right)\right)
\\ &
+P_4\left(\frac{3}{16 \epsilon }-\frac{45}{64}+\frac{63 \epsilon }{64}-\frac{9 \epsilon ^2}{32}+O\left(\epsilon ^4\right) \right).
\end{align}
The three-loop massless propagator master integrals in this relation were evaluated many years ago \cite{IBP}.~\footnote{All the integrals appearing  in \p{C3} can be handled by  {\tt MINCER} \cite{MINCER}. In addition to the single integral \p{C3-np1} contributing to the non-planar correction to the anomalous dimension, we found 76 integrals in the planar sector. Evaluating them with the help of   {\tt MINCER}, we arrived at the same result \p{C3-p}. However, we would like to emphasize that at five loops the reduction of more than 17000 integrals to master integrals can only be done by a direct application of the IBP method (see Sect.~\ref{K5}).}
Using the known results \cite{BC} for the master three-loop integrals $L_1,P_1,\ldots,P_4$ we find from \p{C3-IBP}
\begin{align}\label{C3-p}
C_{3,\rm g=0}  = \frac{39}4 -\frac94 \zeta_3 + \frac{45}8 \zeta_5 + O(\epsilon)\,.
\end{align}
Substituting \p{C3-np} and \p{C3-p} into \p{K-C} we finally obtain the following result
for the four-loop correction to the Konishi anomalous dimension
\begin{align}\label{r}
\gamma_\mathcal{K}^{(4)} =  -\frac{39}4 +\frac94 \zeta_3 - \frac{45}8 \zeta_5 +\frac{r}{N_c^2} \zeta_5\,,
\end{align}
where $r=-{5c_{\rm g=1}^{(4)}/256}$ is an undetermined rational constant. {The planar correction to $\gamma_\mathcal{K}^{(4)}$ is in agreement with the known result \cite{Bajnok:2008bm,Bajnok:2009vm,Arutyunov:2010gb,Fiamberti:2007rj,Fiamberti:2008sh,Velizhanin:2008jd}. The non-planar correction has been computed in \cite{Velizhanin:2009gv}, and the result confirms our prediction. Moreover, it allows us to fix the value of the unknown constant in \p{r}:
\begin{align}\label{}
r = -\frac{135}{2}\,.
\end{align}

In this section we have demonstrated the high efficiency of our method for computing the Konishi anomalous dimension at four loops. We emphasize once again that we do not use the conventional Feynman diagram technique. For comparison, the direct calculation of \cite{Fiamberti:2007rj,Fiamberti:2008sh} involves hundreds of $\cN=1$ super-graphs, each giving rise to a number of Feynman integrals; in the calculation of \cite{Velizhanin:2008jd} the number of contributing Feynman graphs exceeds 130000.}

\section{Konishi anomalous dimension at five loops}\label{K5}

It is straightforward to extend our analysis to five loops. In this case, the correlation
function $F^{(5)}$ has the following form
\begin{align}
F^{(5)}(x_i)= F^{(5)}_{\rm g=0} +\frac1{N_c^2} F^{(5)}_{\rm g=1} +\frac1{N_c^4} F^{(5)}_{\rm g=2} \,.
\end{align}
In what follows, we shall restrict the discussion to the planar sector only. As was shown in Ref.~\cite{Eden:2012tu}, the five-loop correlation function in the planar limit is given by
\begin{align}\label{F5-g=0}
 F^{(5)}_{\rm g=0}(x_i) ={1\over 5!\,(-4\pi^2)^5} \int \frac{  d^4x_5d^4x_6d^4x_7d^4x_8d^4x_9 \, P^{(5)}(x_1,\ldots,x_9) }{x_{56}^2x_{57}^2x_{58}^2x_{59}^2
 x_{67}^2x_{68}^2x_{69}^2x_{78}^2x_{79}^2x_{89}^2
 \prod_{i=1}^4 x_{i5}^2 x_{i6}^2x_{i7}^2x_{i8}^2x_{i9}^2}\,,
\end{align}
where the polynomial $P^{(5)}$ is invariant under  the $S_9$ permutations of the four external
points $x_1,\ldots,x_4$ and the five integration points $x_5,\ldots,x_9$. It is given by
the following expression:
\begin{align}
 P^{\text{(5)}}=& -\ft{1}{2} x_{13}^2 x_{16}^2 x_{18}^2 x_{19}^2 x_{24}^4 x_{26}^2 x_{29}^2 x_{37}^2 x_{38}^2 x_{39}^2 x_{47}^2 x_{48}^2
   x_{56}^2 x_{57}^2 x_{58}^2 x_{59}^2 x_{67}^2  \notag
   \\
 & +\ft{1}{4} x_{13}^2 x_{16}^2 x_{18}^2 x_{19}^2 x_{24}^4 x_{26}^2 x_{29}^2 x_{37}^4 x_{39}^2 x_{48}^4 x_{56}^2 x_{57}^2
   x_{58}^2 x_{59}^2 x_{67}^2 \notag \\
&+\ft{1}{4} x_{13}^4 x_{17}^2 x_{19}^2 x_{24}^2 x_{26}^2 x_{27}^2 x_{29}^2 x_{36}^2 x_{39}^2 x_{48}^6 x_{56}^2 x_{57}^2
   x_{58}^2 x_{59}^2 x_{67}^2\notag  \\
&+\ft{1}{6} x_{13}^2 x_{16}^2 x_{19}^4 x_{24}^4 x_{28}^2 x_{29}^2 x_{37}^4 x_{38}^2 x_{46}^2 x_{47}^2 x_{56}^2 x_{57}^2
   x_{58}^2 x_{59}^2 x_{68}^2\notag \\
 &-\ft{1}{8} x_{13}^4 x_{16}^2 x_{18}^2 x_{24}^4 x_{28}^2 x_{29}^2 x_{37}^2 x_{39}^2 x_{46}^2 x_{47}^2 x_{56}^2 x_{57}^2
   x_{58}^2 x_{59}^2 x_{69}^2 x_{78}^2\notag \\
& +\ft{1}{28} x_{13}^2 x_{17}^2 x_{18}^2 x_{19}^2 x_{24}^8 x_{36}^2 x_{38}^2 x_{39}^2 x_{56}^2 x_{57}^2 x_{58}^2 x_{59}^2
   x_{67}^2 x_{69}^2 x_{78}^2 \notag \\
 &+\ft{1}{12} x_{13}^2 x_{16}^2 x_{17}^2 x_{19}^2 x_{26}^2 x_{27}^2 x_{28}^2 x_{29}^2 x_{35}^2 x_{38}^2 x_{39}^2 x_{45}^2
   x_{46}^2 x_{47}^2 x_{49}^2 x_{57}^2 x_{58}^2 x_{68}^2+\text{$S_9$ permutations} \,,
\label{P5}
\end{align}
where the relative coefficients
follow from the requirement for the correlation function to have the correct asymptotic
behaviour at short distances.

The analysis goes along the same lines as at four loops. We start with examining the
integrand \p{F5-g=0} in the double short-distance limit $x_2\to x_1$ and $x_4\to x_3$
and, then, apply \p{Is} to identify the five-loop integrand of $I^{(5)}$:
\begin{align}\label{P4-hat}
\notag
\mathcal{I}_5 =  -\frac{6}{5! (4\pi^2)^5}  \frac{x_{13}^4}{\prod_{i=5}^9 x_{1i}^4 x_{3i}^4}
\bigg[&
\frac1{5!} \frac{\widehat{P}_{5,6,7,8,9}}{x_{56}^2x_{57}^2x_{58}^2x_{59}^2
 x_{67}^2x_{68}^2x_{69}^2x_{78}^2x_{79}^2x_{89}^2} -\frac{1}{4}x_{12}^4\frac{\widehat{P}_{5,6,7,8}}{x_{56}^2x_{57}^2x_{58}^2 x_{67}^2 x_{68}^2x_{78}^2}
 \\ \notag
& - \frac12 x_{13}^4 \frac{  \widehat{P}_{5,6,7}}{x_{56}^2 x_{57}^2 x_{67}^2}\frac{ \widehat{P}_{8,9}}{x_{89}^2} + 6 (x_{13}^4)^2\frac{\widehat{P}_{5,6,7}}{x_{56}^2 x_{57}^2 x_{67}^2}
+9 (x_{13}^4)^2\frac{ \widehat{P}_{5,6}}{x_{56}^2}\frac{ \widehat{P}_{7,8}}{x_{78}^2}
\\
& - 108 (x_{13}^4)^3\frac{ \widehat{P}_{5,6}}{x_{56}^2} + \frac{1296}{5}(x_{13}^4)^4\bigg]+
\text{$S_5$ permutations}\,,
\end{align}
where $\widehat{P}_{5,6,7,8,9}$ is the polynomial $P^{(5)}$ evaluated at  $x_2=x_1$ and $x_4= x_3$ and the remaining $\widehat{P}-$polynomials were defined
earlier (see Eqs.~\p{P2-hat}, \p{P3-hat} and \p{I4-p}). The expression on the right-hand side of \p{P4-hat} is symmetrized with respect to $S_5$ permutations of the integration points
$x_5,\ldots,x_9$. To save space, here we do not present the explicit expression for $\mathcal{I}_5$.

At the next step, we simplify the expression for $\mathcal{I}_5$ by choosing all integration
points $x_5,\ldots,x_9$ to lie in the vicinity of the point $x_1$. This is equivalent to
sending the external point to infinity $x_3\to \infty$ with all remaining points fixed
\begin{align}
 \widehat{\mathcal{I}}_{5} = \lim _{x_3\to\infty}
 \mathcal{I}_{5} (x_1,x_3; x_5,\ldots,x_9)\,.
\end{align}
Finally, we apply \p{K-C} to express the coefficient $C_4$ as the following four-loop integral
 \begin{align} \label{C4}
C_4= \frac{2\pi^2}{(x_{15}^2)^{2+4\epsilon}} \int d^{4-2\epsilon} x_6 d^{4-2\epsilon} x_7 d^{4-2\epsilon} x_{8}d^{4-2\epsilon} x_{9}\,
\widehat{\mathcal{I}}_{5}  (x_1; x_5,\ldots,x_9) \,.
\end{align}
As before, to simplify the calculation we put $x_5=0$.

\begin{figure}[t]
\psfrag{M44}[cc][cc]{$M_{44}$}\psfrag{M61}[cc][cc]{$M_{61}$}\psfrag{M36}[cc][cc]{$M_{36}$}\psfrag{M31}[cc][cc]{$M_{31}$}\psfrag{M35}[cc][cc]{$M_{35}$}
\psfrag{M22}[cc][cc]{$M_{22}$}\psfrag{M32}[cc][cc]{$M_{32}$}\psfrag{M33}[cc][cc]{$M_{33}$}\psfrag{M34}[cc][cc]{$M_{34}$}\psfrag{M25}[cc][cc]{$M_{25}$}
\psfrag{M23}[cc][cc]{$M_{23}$}\psfrag{M27}[cc][cc]{$M_{27}$}\psfrag{M24}[cc][cc]{$M_{24}$}\psfrag{M26}[cc][cc]{$M_{26}$}\psfrag{M01}[cc][cc]{$M_{01}$}
\psfrag{M21}[cc][cc]{$M_{21}$}\psfrag{M12}[cc][cc]{$M_{12}$}\psfrag{M11}[cc][cc]{$M_{11}$}\psfrag{M14}[cc][cc]{$M_{14}$}\psfrag{M13}[cc][cc]{$M_{13}$}
\centerline{\includegraphics[width=0.95\linewidth]{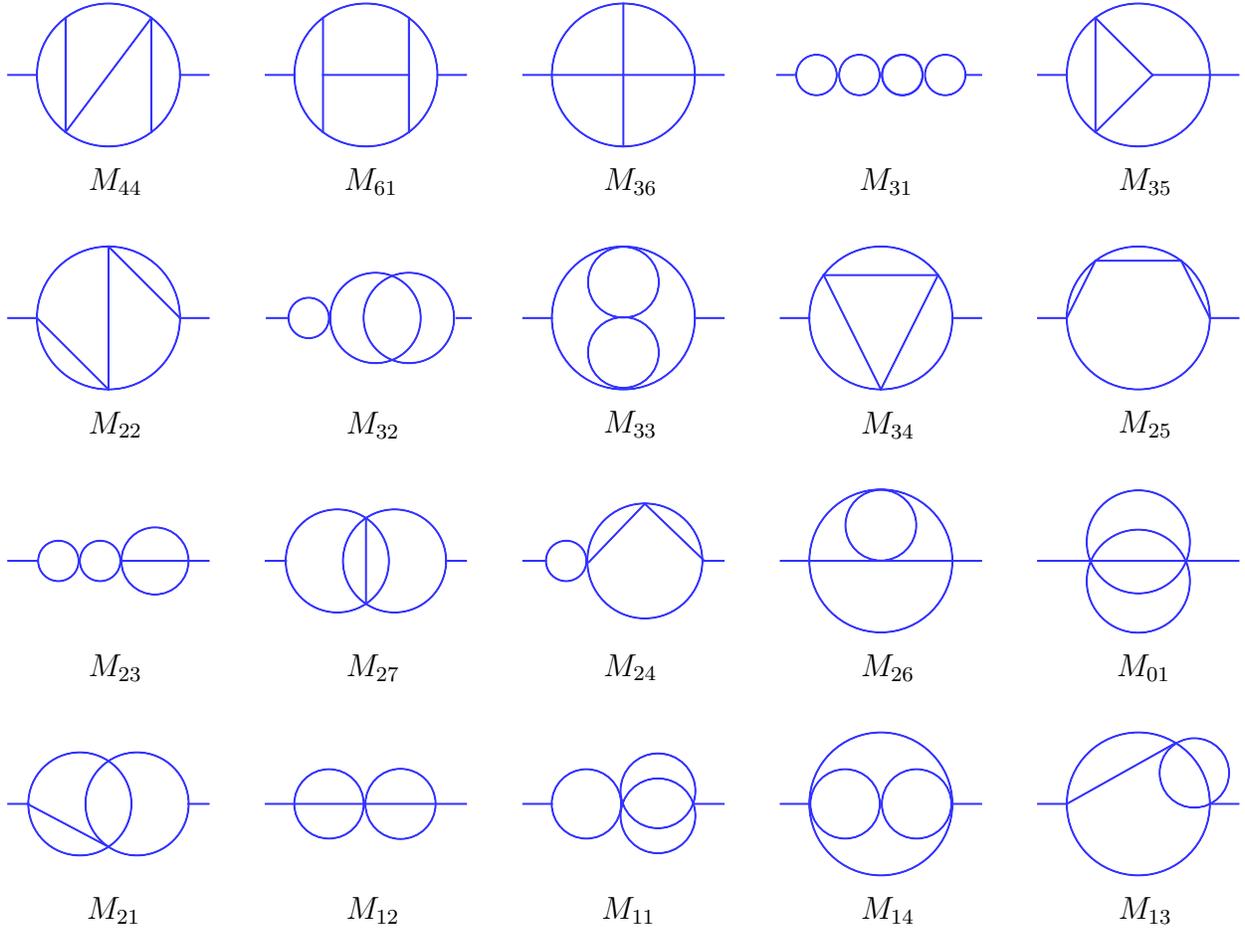}}
\caption{Diagrammatic representation of the planar basis integrals in the dual momentum representation. Blue line denote momentum propagators
$1/k^2$ with the momentum $k=x_i-x_j$.}
\label{fig-basis}
\end{figure}

Replacing   $\widehat{\mathcal{I}_{5}}$ in \p{C4} by its explicit expression, we find that
$C_4$ is given by the sum of more than 17000 four-loop two-point Feynman integrals.
All of them belong to the following family of four-fold
integrals, with various integer (positive and negative) indices $a_1,\ldots,a_{14}$
\begin{align} \notag
G(a_1,\ldots,a_{14})= \int &
\frac{d^{4-2\epsilon} x_6 d^{4-2\epsilon} x_7 d^{4-2\epsilon} x_8d^{4-2\epsilon} x_9}{
 (x_{16}^2)^{a_1} (x_{17}^2)^{a_2}(x_{18}^2)^{a_3}(x_{19}^2)^{a_4}
(x_{6}^2)^{a_5} (x_{7}^2)^{a_6}(x_{8}^2)^{a_7}}
\\
& \times
\frac{1}{(x_{9}^2)^{a_8}  (x_{67}^2)^{a_9} (x_{68}^2)^{a_{10}}(x_{69}^2)^{a_{11}} (x_{78}^2)^{a_{12}}(x_{79}^2)^{a_{13}} (x_{89}^2)^{a_{14}}}\,.
\end{align}
As in the four-loop case, to evaluate each of them we apply the IBP method \cite{IBP}. To solve the IBP relations, i.e.
to represent every integral on the right-hand side of \p{C4} as a linear combination of master integrals,
we apply the {\tt C++} version of the code {\tt FIRE} \cite{FIRE}. In this way, we found that
$C_4$ is given by a linear combination of 22 master integrals:
\begin{align} \label{C4-master} \notag
C_4& =  w_{44} M_{44} + w_{61} M_{61} + w_{36} M_{36} + w_{31} M_{31} + w_{35} M_{35} + w_{22} M_{22} + w_{32} M_{32}
\\ \notag
& + w_{33} M_{33} +  w_{34} M_{34} + w_{25} M_{25} + w_{23} M_{23} + w_{27} M_{27} + w_{24} M_{24} + w_{26} M_{26}
\\
& + w_{01} M_{01} +
w_{21} M_{21}+  w_{12} M_{12} + w_{11} M_{11} + w_{14} M_{14} + w_{13} M_{13} + w_1 I_1 + w_2 I_2\,,
\end{align}
with the coefficient functions $w_i$ defined below in Eq.~\p{w}. Among the master
integrals only two, $I_1$ and $I_2$, are associated with
non-planar graphs (see Eqs.~\p{master-np} below). The remaining 20 master integrals
$M_{44},\ldots,M_{13}$ correspond to planar graphs. This allows us
to introduce the dual
momenta $k_i=x_i-x_{i+1}$ and represent the same integrals as four-loop propagator master (momentum)
integrals shown in Fig.~\ref{fig-basis}. The latter integrals were calculated
recently in \cite{BC} as an $\epsilon$ expansion up to transcendentality weight 
seven.~\footnote{At the moment, results for the master integrals are known
up to transcendentality weight twelve \cite{LSS}.} The explicit expressions for the planar
integrals $M_{44},\ldots,M_{13}$ can be found in \cite{BC}. To save space, we do not
present them here.

The corresponding
coefficient functions are given by
\begin{align} \label{w} \notag
w_{44} &=
   -\ft{3}{80}+\dots, 
\\ \notag
w_{61}& =\ft{3}{128}+\ft{3 }{32} \epsilon+\dots, 
\\ \notag
w_{36}&=
   \ft{1}{10}\epsilon ^{-1} -\ft{189}{320}+\ft{11}{16} \epsilon+\dots, 
\\ \notag
w_{31}& =-\ft{3}{64 }\epsilon ^{-2}+\ft{81}{320
   }\epsilon ^{-1}-\ft{27}{40}+\ft{57}{80}\epsilon -\ft{9}{40} \epsilon ^2+\dots, 
\\ \notag
 w_{35} &=   -\ft{3}{80}\epsilon ^{-2} +\ft{233}{480}\epsilon ^{-1}-\ft{189}{80}+\ft{2131 }{480}\epsilon -\ft{509}{80} \epsilon ^2+\dots, 
\\ \notag
w_{22}&=     -\ft{11}{20
 }  \epsilon ^{-3}+\ft{295}{64} \epsilon ^{-2}-\ft{85}{32 } \epsilon^{-1}-\ft{39711}{640}+\ft{349167
   }{1280} \epsilon-\ft{358191 }{512}\epsilon ^2+\ft{1748923}{1024} \epsilon ^3+\dots, 
\\ \notag
w_{32}&=
   -\ft{3}{80} \epsilon ^{-3}+\ft{9}{32}\epsilon ^{-2} -\ft{261}{160}\epsilon ^{-1}+\ft{1281}{160}-\ft{1953}{64}\epsilon+\ft{3879}{40}\epsilon ^2-\ft{2313}{8} \epsilon
   ^3+\ft{1701}{2} \epsilon ^4+\dots, 
\\ \notag
   w_{33}&=     -\ft{7}{80} \epsilon
   ^{-3}+\ft{57}{64} \epsilon ^{-2}-\ft{179}{40 } \epsilon^{-1}+\ft{4911}{320}-\ft{6269 }{160}\epsilon +\ft{1723
 }{20}  \epsilon ^2-\ft{2631}{16} \epsilon ^3+\ft{17481}{80} \epsilon ^4+\dots, 
\\ \notag
w_{34}&=
   -\ft{27}{320} \epsilon ^{-3}+\ft{21}{20} \epsilon ^{-2}-\ft{3629}{640
  } \epsilon^{-1} +\ft{1141}{64}-\ft{5643}{128} \epsilon +\ft{475}{4} \epsilon ^2-\ft{25953 }{80}\epsilon
   ^3+\ft{8817}{10} \epsilon ^4+\dots, 
\\ \notag
w_{25}&=
   \ft{9}{80} \epsilon ^{-4}-\ft{313}{160} \epsilon ^{-3}+\ft{4221}{320} \epsilon
   ^{-2}-\ft{25583}{640 } \epsilon^{-1}+\ft{42073}{640}-\ft{3341 }{32} \epsilon+\ft{12279}{64} \epsilon ^2-\ft{4131
  }{32} \epsilon ^3+\dots, 
\\ \notag
w_{23 }&=     -\ft{3}{80
}   \epsilon ^{-4}-\ft{3}{10} \epsilon ^{-3}+\ft{81}{80} \epsilon ^{-2}+\ft{9939}{640  }\epsilon^{-1}
  -\ft{83889}{640}+\ft{183273}{320} \epsilon -\ft{622629}{320} \epsilon ^2+\ft{978141}{160} \epsilon
   ^3+\dots, 
\\ \notag
w_{27}&=
   \ft{21}{160} \epsilon ^{-4}-\ft{753}{320} \epsilon
   ^{-3}+\ft{5319}{320} \epsilon ^{-2}-\ft{9963}{160} \epsilon^{-1} +\ft{27913}{160}-\ft{42339
   }{80} \epsilon+\ft{139719}{80} \epsilon ^2-\ft{448299}{80} \epsilon ^3+\dots, 
\\ \notag
w_{24 }&=
   -\ft{3}{20} \epsilon ^{-4}+\ft{439}{160
 }  \epsilon ^{-3}-\ft{5141}{320} \epsilon ^{-2}+\ft{20549}{640} \epsilon^{-1} +\ft{16623}{1280}-\ft{630099
   }{2560} \epsilon+\ft{5544039}{5120} \epsilon ^2-\ft{8490959 }{2048}\epsilon ^3+\dots, 
\\ \notag
w_{26 }&=
   \ft{19}{80}
   \epsilon ^{-4}-\ft{1329}{320} \epsilon ^{-3}+\ft{127}{4} \epsilon ^{-2}-\ft{43443}{320 }\epsilon^{-1}
   +\ft{504477}{1280}-\ft{2516491 }{2560} \epsilon+\ft{10666651}{5120} \epsilon ^2-\ft{22997991 }{10240}\epsilon
   ^3+\dots, 
\\ \notag
w_{01 }&=
   -\ft{213}{4} \epsilon ^{-5}+\ft{25311}{40} \epsilon ^{-4}-\ft{1038647}{960} \epsilon ^{-3}-\ft{8046763}{480
}   \epsilon ^{-2}+\ft{38217863}{320} \epsilon^{-1} -\ft{398050007}{960}+\ft{2071600273 }{1920}\epsilon +\dots, 
\\ \notag
w_{21 }&=
   -\ft{133}{160} \epsilon ^{-4}+\ft{61}{6} \epsilon ^{-3}-\ft{30023}{480} \epsilon
   ^{-2}+\ft{84321}{320} \epsilon^{-1} -\ft{329747}{384}+\ft{9067319 }{3840} \epsilon-\ft{14362073}{2560} \epsilon
   ^2+\ft{59375477}{5120} \epsilon ^3+\dots, 
\\ \notag
w_{12 }&=
   -\ft{61}{64} \epsilon ^{-5}+\ft{2149}{160} \epsilon
   ^{-4}-\ft{122841}{1280} \epsilon ^{-3}+\ft{576611}{1280} \epsilon ^{-2}-\ft{2066537}{1280  }\epsilon^{-1}
  +\ft{6439581}{1280}-\ft{1803913}{128} \epsilon +\ft{11613069}{320} \epsilon ^2+\dots, 
\\ \notag
w_{11 }&=
   -\ft{471}{160} \epsilon ^{-5}+\ft{15583}{320} \epsilon ^{-4}-\ft{346883}{960
  } \epsilon ^{-3}+\ft{509833}{320} \epsilon ^{-2}-\ft{5155033}{960 } \epsilon^{-1}+\ft{1322649}{80}-\ft{1855317
   }{40} \epsilon+\ft{9142761}{80} \epsilon ^2+\dots, 
\\ \notag
w_{14 }&=
   -\ft{51}{160} \epsilon ^{-5}+\ft{361}{64} \epsilon ^{-4}-\ft{4829}{80}\epsilon ^{-3}+\ft{471857}{960}  \epsilon ^{-2}-\ft{677747}{240} \epsilon^{-1} +\ft{14881773}{1280}-\ft{289328273}{7680} \epsilon +\ft{1662945973
 }{15360}  \epsilon ^2+\dots, 
\\
w_{13 }&=
   \ft{573}{160} \epsilon ^{-5}-\ft{5431}{80} \epsilon
   ^{-4}+\ft{1084379}{1920} \epsilon ^{-3}-\ft{1648533}{640} \epsilon ^{-2}+\ft{16101181}{1920
   } \epsilon^{-1}-\ft{16307253}{640}+\ft{12145053 }{160} \epsilon-\ft{65454483 }{320}\epsilon ^2+\dots
\end{align}
Here the series expansion of $w_i$ is truncated at the order in $\epsilon$ which is related to the maximal power of $1/\epsilon$ in the expression for the corresponding basis  integral $M_i$ in \p{C4-master}, so that the right-hand side of \p{C4-master} can be evaluated at order $O(\epsilon^0)$.

\begin{figure}[t]
\psfrag{x1}[cc][cc]{$\scriptstyle x_{1}$}
\psfrag{x5}[cc][cc]{$\scriptstyle 0$}
\psfrag{x6}[cc][cc]{$\scriptstyle x_{6}$}
\psfrag{x7}[cc][cc]{$\scriptstyle x_{7}$}
\psfrag{x8}[cc][cc]{$\scriptstyle x_{8}$}
\psfrag{x9}[cc][cc]{$\scriptstyle x_{9}$}
\psfrag{2}[cc][cc]{$\scriptstyle 2$}
\psfrag{I1}[cc][cc]{$I_{1}$}\psfrag{I2}[cc][cc]{$I_{2}$}\psfrag{I3}[cc][cc]{$I_{3}(0)$}\psfrag{I4}[cc][cc]{$I_{4}(0)$}
\centerline{\includegraphics[width=0.95\linewidth]{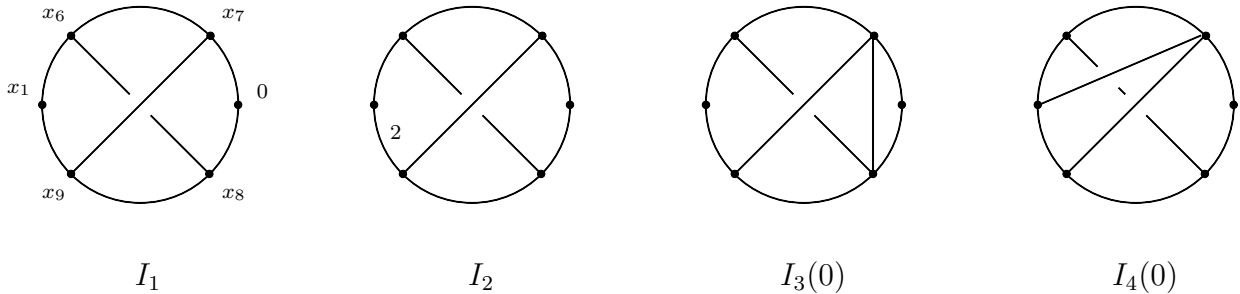}}
\caption{Diagrammatic representation of the non-planar basis integrals $I_1$, $I_2$ and two auxiliary integrals $I_3(0)$,
$I_4(0)$ defined in Eqs.~\p{master-np} and \p{aux}, respectively. The line with the index 2 denotes a square of scalar propagator $1/(x^2)^2$, while all the remaining lines stand for $1/x^2$. The points $x_1$ and $0$ are external and integration goes over the points $x_6,\ldots,x_9$.}
\label{fig-basis-np}
\end{figure}

The two non-planar master integrals $I_1$ and $I_2$ entering the right-hand side of \p{C4-master} are shown diagrammatically in Fig.~\ref{fig-basis-np} and their explicit form can be found in \p{master-np}. These integrals are evaluated in Appendix~\ref{app-master} leading to
\begin{align} \label{I-np} \notag
I_1 &=\frac{5\,\zeta_5}{\epsilon} + \frac{5}{378} \pi^6 - 13\, \zeta_3{}^2 +
  35 \,\zeta_5+\lr{-\frac{13}{30}\pi ^4 \zeta_3-91 \zeta_3{}^2+195 \zeta_5 -\frac{5 }{3}\pi ^2
   \zeta_5+\frac{345 }{4}\zeta_7+\frac{5 }{54} \pi ^6}\epsilon+\ldots
\\
I_2 &=-\frac{20\,\zeta_5}{\epsilon} -\frac{10}{189} \pi^6 - 8\, \zeta_3{}^2 -
  40\, \zeta_5+\lr{ -\frac{4}{15}\pi ^4 \zeta_3-16 \zeta_3{}^2-80 \zeta_5+\frac{20 }{3}\pi ^2 \zeta_5-520 \zeta_7-\frac{20}{189} \pi ^6}\epsilon+\ldots
\end{align}
Their coefficient functions in \p{C4-master} are
\begin{align}\label{w12} \notag
w_1 &=\frac{3}{80} \epsilon^{-1} -\frac{21}{80}+\frac{741}{640} \epsilon +O\left(\epsilon ^2\right),
\\[2mm]
w_2 &= \frac{9}{160
   } \epsilon^{-1} -\frac{9}{80}+\frac{807}{320} \epsilon +O\left(\epsilon ^2\right)\,.
\end{align}

Finally, we combine together the relations \p{w} -- \p{w12},
make use of the results of Ref.~\cite{BC} for the master integrals shown in Fig.~\ref{fig-basis} and obtain from \p{C4-master}  the
following result for $C_4$ or equivalently, five-loop Konishi anomalous dimension
\begin{align}\label{main}
\gamma_{\mathcal{K}}^{(5)}=-C_4=\frac{237}{16}+\frac{27}{4} \zeta_3-\frac{81}{16} \zeta _3{}^2-\frac{135}{16} \zeta_ 5+\frac{945}{32}\zeta_7\,.
\end{align}
This relation is the main result of the paper. It is in perfect agreement with the prediction of the integrable models \cite{Bajnok:2008bm,Bajnok:2009vm,Arutyunov:2010gb,Balog:2010xa}.

\section{Conclusions}

In this paper we have developed a new efficient method for the computation of the Konishi
anomalous dimension at higher loops. It does not use the conventional Feynman diagram
technique with the associated very large number of contributing graphs and Feynman
integrals. Instead, we exploited the recently discovered new symmetry of the four-point
correlation function of $\cN=4$ SYM stress-tensor multiplets to predict the form of its integrand 
as a linear combination of a small number of relevant diagrams. Then, we examined the
asymptotic behaviour of the logarithm of the four-point correlation function in the double
short-distance limit and related the Konishi anomalous dimension to its leading
logarithmic singularity.  Finally, by analyzing the expected singularity of the logarithm of the correlation function in this limit, we were able to lower the loop order of the  contributing Feynman integrals by one, that is to express the Konishi anomalous dimension at $\ell$ loops in terms of  finite two-point integrals at $(\ell-1)$ loops.
Going through these steps, we obtained the five-loop Konishi anomalous dimension in the
planar limit as a sum of 22 master four-loop two-point integrals. Replacing the master
integrals by their explicit expressions we arrived at an analytic result for this anomalous dimension
which agrees with the integrability prediction \cite{Bajnok:2008bm,Bajnok:2009vm,Arutyunov:2010gb,Balog:2010xa}.

At present, the expression for the integrand of the four-point correlation function is known
up to six loops in the planar limit \cite{Eden:2012tu}. In our calculation of the Konishi anomalous dimension we made use of this expression to five loops only.
By applying the method developed in this paper, it is straightforward to extend the
analysis to six loops and to express the six-loop Konishi anomalous
dimension in terms of five-loop two-point integrals. 
The evaluation of such integrals is still an open problem, not because of
their number, but because the IBP reduction to master integrals is
a very complicated problem at this level. Still, we are optimistic that
further development in  this  direction will eventually make the six-loop calculation possible.
Likewise, no prediction for the six-loop Konishi anomalous dimension is available from AdS/CFT considerations, and to obtain it using the existing integrability approaches appears to be a rather non-trivial task.

Another result of our study is the prediction of the non-planar correction to the
Konishi anomalous dimension at four loops in the form $r\zeta_5/N_c^2$ with $r$ being an
undetermined rational number. Our prediction is in full agreement with the result of the
direct Feynman diagram calculation in \cite{Velizhanin:2009gv}, which also allows us to fix
the value of $r=-135/2$. It is interesting to note that in our approach the non-planar
correction at four loops originates from just a single and very simple three-loop propagator
integral shown in Fig.~\ref{fig-np}.  We would like to emphasize that the non-planar
$O(1/N_c^2)$ correction to the four-point correlation function derived in Ref.~\cite{Eden:2012tu}
depends on four arbitrary rational constants. The parameter $r$ is given by  a particular linear combination of these coefficients. To fix each of the four coefficients we need three more relations. They can be obtained from the comparison of the non-planar corrections
to the twist-two anomalous dimensions computed in Ref.~\cite{Velizhanin:2009gv} with the analogous results
 from the OPE analysis of the non-planar correction to the four-point correlation function.
As explained in Ref.~\cite{Eden:2012tu}, the perturbative corrections to the correlation function
have an iterative structure at higher loops. In application to the Konishi operator, this implies that the non-planar $O(1/N_c^2)$ correction to its anomalous dimension is uniquely
defined at all loops by the values of these four coefficients. We would like to mention
that starting from five loops, the anomalous dimension receives $O(1/N_c^4)$ corrections.
The method proposed in this paper can be equally applied to the study of such corrections.

The Konishi operator is just the first in an infinite series of twist-two operators, all appearing in the OPE of two $\cN=4$ SYM stress-tensor multiplets. The four-point function that we use for the evaluation of $\gamma_\cK$ contains the information about the whole spectrum
of anomalous dimensions
of twist-two (as well as higher twist) operators. However, in order to
extract it from the OPE, one needs to either evaluate analytically all
relevant higher-loop four-point conformal integrals, or at least to
work out their asymptotic expansion in the double short-distance
limit. This problem is not yet solved in full generality beyond two
loops\footnote{Although at three loops, the inverse process has been
  partially done,
  namely a prediction has been made for the three-loop correlation
  function in the limit
  $u\rightarrow 0$ but with finite $v$~\cite{hughfrancispaul}, by
  making use of
  the three-loop twist-two, arbitrary spin anomalous dimensions predicted
  in~\cite{Kotikov:2004er} together with known lower-loop twist-two
  data, and
  a conformal partial wave analysis. See
  appendix~\ref{sec:twist-two-anomalous} for more on higher spin
  twist-two anomalous dimensions and the four-point correlation function.}, but it undoubtedly deserves further attention.

\section*{Acknowledgements}

We are grateful to Henrik Johansson  for a
number of enlightening discussions, to Kostya Chetyrkin for useful correspondence and to Alexander Smirnov for the possibility
to use his C++ version of {\tt FIRE}. G.K. and E.S. acknowledge partial support by the French National Agency
for Research (ANR) under contract StrongInt
(BLANC-SIMI-4-2011). B.E. is supported by the Deutsche Forschungsgemeinschaft (DFG), Sachbeihilfe ED 78/4-1. P.H. acknowledges support from an
STFC Consolidated Grant number ST/J000426/1. V.S. acknowledges support from the RFBR through grant
11-02-01196 and from DFG through SFB/TR~9 ``Computational Particle Physics''.

\appendix

\makeatletter
\def\@seccntformat#1{Appendix\ \csname the#1\endcsname\quad}
\makeatother

\section{IR rearrangement in coordinate space}

In this appendix we explain in detail the method that we employ in our calculation of
the Konishi anomalous dimension. It represents an extension
of the so-called infrared rearrangement method (IRR) \cite{IRR} to coordinate
space.

To describe the method, let us consider as an example the following  four-loop integral
in Euclidean $D-$dimensional space-time (with $D=4-2\epsilon$)
\begin{align}
I(x_{13})= \frac{\e^{4\gamma  \epsilon}}{\pi^{2D}}  \int
\frac{(x_{13}^2)^4\,d^D x_5 \ldots  d^D x_8}{
x_{15}^2 x_{16}^2 x_{17}^2 x_{18}^2 x_{35}^2 x_{36}^2 x_{37}^2 x_{38}^2 x_{56}^2 x_{68}^2 x_{78}^2 x_{57}^2}
\,.
\label{int12}
\end{align}
We would like to stress that $x_i$ are true coordinates in configuration space and, therefore, $I(x_{13})$ is different from the conventional integrals that one encounters in dimensional regularization in which case all distances in the denominator appear with power $(1-\epsilon)$.

The integral \p{int12} has a simple pole in $\epsilon$
\begin{align}\label{I-C}
 I(x_{13}) =(x_{13}^2)^{-4\epsilon}\left[ \frac{C }{\epsilon} + O(\epsilon^0) \right]\,.
\end{align}
It comes from integration over
the region where $x_5,\ldots,x_8$ are all close to $x_1$ and from the
symmetrical region where $x_5,\ldots,x_8$ are all close to $x_3$.
Since the integration variables are true coordinates in Euclidean space,
the pole $1/\epsilon$ has to be interpreted as an UV divergence.
Notice that the integrand of $I(x_{13})$ coincides (up to an overall normalization factor) with
\p{I4-np}. As a result, the residue $C$
defines the four-loop non-planar correction to the Konishi anomalous dimension.

In general, the UV divergences in coordinate space come from regions where
the integrand considered as a generalized function of $x_i$
(tempered distribution, i.e. linear functional on a space of test functions)
is ill-defined. In our example, the product of $x^2-$factors in the denominator of \p{int12}
turns out to be unintegrable in a vicinity of the two external points, $x_1$ and $x_3$. In the first
case, we consider the product
\begin{align}
F(x_1,x_5,\ldots,x_8)= \frac{1}{
x_{15}^2 x_{16}^2 x_{17}^2 x_{18}^2 x_{56}^2 x_{68}^2 x_{78}^2 x_{57}^2}
\label{int12TD}
\end{align}
as a tempered distribution. Its divergent part is described by an
UV counter-term
\footnote{In this example, $\Delta$  contains a simple pole in $\epsilon$. In a general situation, this would be a finite linear combination of negative powers of $\epsilon$.}
\begin{align}
\Delta(x_1,x_5,\ldots,x_8)=\frac{C}{2\epsilon} \delta(x_1-x_5)\ldots \delta(x_1-x_8)\,,
\label{int12ppCT}
\end{align}
with the constant $C$ determined below. Similar counter-term $\Delta(x_3,x_5,\ldots,x_8)$
describes singular behaviour of the integrand \p{int12} in the vicinity of $x_3$.
Thus, the pole part of (\ref{int12}) is  just twice the factor $C/(2\epsilon)$ in
(\ref{int12ppCT})
\begin{align}
I= \int
 d^D x_5 \ldots  d^D x_8\left[  \Delta(x_1,x_5,\ldots,x_8)+\Delta(x_3,x_5,\ldots,x_8)\right]  + O(\epsilon^0) = \frac{C}{\epsilon}+ O(\epsilon^0) \,,
\label{int12pp}
\end{align}
leading to \p{I-C}.

To evaluate the constant $C$ in \p{int12ppCT} we apply the infrared rearrangement (IRR) method
originally proposed by Vladimirov in Ref.~\cite{IRR} in momentum space. It makes use of the fact that, for
an infrared finite but logarithmically UV-divergent Feynman integral without subdivergences,
the contribution of the counter-term is just a constant. The idea of IRR is
to set the external momenta to zero and then, in order to avoid the appearance of
IR divergences, to introduce an external momentum (or a mass) in such a way that
the calculation becomes simpler.\footnote{If it is not possible to avoid such IR divergences
one can remove them immediately by the so-called $R^*$-operation \cite{RStar1,RStar2} but
we do not meet such a complication in our calculations.}

Applying the IRR method to \p{int12}, we should have transformed the integral $I(x_{13})$ to
momentum space  via Fourier transform. However we will not do this for the following two reasons.
First, the resulting momentum integral will be a four-loop one while we can obtain the residue $C$  from a three-fold integral only as described below.
It is well known \cite{RStar2}  that the evaluation of the UV pole part of a given $\ell-$loop
momentum-space Feynman integral can be reduced to evaluating massless propagator
$(\ell-1)-$loop Feynman integrals to order $\epsilon^0$. However, as was already mentioned,
the integral \p{int12} is different from the conventional Feynman integral. In particular,
the  $1/x^2-$factors on the right-hand side of  \p{int12}
are replaced in the momentum representation by the factors of $1/(k^2)^{1-\epsilon}$ depending on $\epsilon$. These are much more complicated objects, both from the point
of view of an IBP reduction and evaluating master integrals, so that it is
the second reason why we want to stay in coordinate space.

Let us apply the IRR method to \p{int12TD} in coordinate space and treat the coordinates
$x_1,x_5$ as external and $x_6,x_7,x_8$ as internal points. Notice that
setting an external momentum to zero corresponds to integrating over the corresponding coordinate. Then, the constant $C$ in (\ref{int12ppCT}) can be obtained by integrating
both sides of \p{int12TD} with respect to internal points
\begin{align}
F(x_1,x_5)=\int  \frac{ d^D x_5 d^D x_6 d^D x_7}{
x_{15}^2 x_{16}^2 x_{17}^2 x_{18}^2
x_{56}^2 x_{68}^2 x_{78}^2 x_{57}^2}= \frac{C}{2\epsilon}\delta(x_1-x_5) + O(\epsilon^0)\,.
\label{F18a}
\end{align}
The integral on the left-hand side depends on the two external points and is of propagator
type. We can check it has no IR divergences, i.e. divergences at large values
of coordinates, and has the following form by dimensional arguments
\begin{align}
F(x_1,x_5)= f(\epsilon)\frac{1}{(x_{15}^2)^{2+3\epsilon}}\,.
\label{F18}
\end{align}
Here the only source of the simple pole in $\epsilon$ is hidden in the second factor (which
is considered as a distribution) so that $f(\epsilon)$ is analytic in a vicinity
of the point $\epsilon=0$. The simplest way to reveal the $1/\epsilon$ pole of the distribution
$1/(x_{15}^2)^{2+3\epsilon}$  is to take its $D$-dimensional Fourier transform with a help of the identity
\begin{align}
{\cal F} \left[\frac{1}{(x^2)^\lm}\right]
=\frac1{\pi^{D/2}
}\int  d^D x\, \e^{i p x}\frac{1}{(x^2)^\lm}
=\frac{4^{D/2-\lm}}{\Gm(\lm)}
\frac{\Gm(D/2-\lm)}{(p^2)^{D/2-\lm}}\,.
\label{Fourier}
\end{align}
In particular, for $\lm=2+3\epsilon$ we find from \p{F18} (for $x_5=0$)
\bea
{\cal F} \left[F(x_1,0)\right]
= f(\epsilon) \frac{4^{-4\epsilon}\Gm(-4\epsilon)}{\Gm(2+3\epsilon)}
\frac{1}{(p^2)^{-4\epsilon}} = -\frac{f(0)}{4\epsilon} + O(\epsilon^0)\,.
\eea
At the same time, replacing $F(x_1,0)$ by its expression  \p{F18a} we
obtain the left-hand side of this relation as $C/(2\epsilon) + O(\epsilon^0)$ leading to
\begin{align}
C =  -\frac12 f(0) = -\frac12 F(x_1,0)\bigg|_{x_{1}^2=1, D=4}\,.
\end{align}
It is easy to see that the integral $F(x_1,x_5)$, Eq.~\p{F18a}, corresponds to
a planar graph shown in Fig.~\ref{fig-np}. After going to the dual momenta,
we find that it coincides with the graph $N_{2}$ of Baikov and Chetyrkin \cite{BC} leading to
\begin{align}
C=-10\,\zeta(5)\,.
\end{align}

\section{Non-planar master integrals} \label{app-master}

In this appendix, we evaluate the two non-planar master Euclidean integrals \p{I-np}.
They  have the following form (with $D=4-2\epsilon$ and $x_1^2=1$)
\begin{align}\notag
I_1 &= \frac{\e^{4\gamma  \epsilon}}{\pi^{2D}} \int \frac{d^D x_6 d^D x_7 d^D x_8 d^D x_9}{x_{16}^2 x_{19}^2 x_{67}^2 x_{68}^2x_{7}^2x_{79}^2 x_{8}^2 x_{89}^2} = \frac{a_1}{\epsilon} + b_ 1+ c_1\epsilon + O(\epsilon^2)\,,
\\ \label{master-np}
I_2 &=\frac{\e^{4\gamma  \epsilon}}{\pi^{2D}} \int \frac{d^D x_6 d^D x_7 d^D x_8 d^D x_9}{x_{16}^2 (x_{19}^2)^2 x_{67}^2 x_{68}^2x_{7}^2x_{79}^2 x_{8}^2 x_{89}^2} = \frac{a_2}{\epsilon} + b_2+ c_2\epsilon+ O(\epsilon^2)\,.
\end{align}
Here we introduced the factor in front of the integrals to avoid the appearance of terms proportional to $\ln\pi$
and Euler's constant $\gamma$ in
the right-hand side.
The diagrammatic representation of $I_1$ and $I_2$ is shown in Fig.~\ref{fig-basis-np}. Both
integrals develop poles $1/\epsilon$ but their origin is different. For the integral $I_1$
it comes from integration over $x_6,x_7,x_8,x_9$ going to infinity simultaneously and,
therefore, has an IR origin. For the integral $I_2$ the pole  comes from integration at
short distances $x_{19}\to 0$ and has UV origin.~\footnote{Since the Euclidean integrals in \p{master-np} are positive definite,
this explain why their residues at the pole, $a_1$ and $a_2$, have opposite signs (see Eqs.~\p{a1} and \p{a2} below).}

Substituting  Eqs.\p{w} -- \p{w12} and \p{master-np} into \p{C4-master}  and
making use of the results of Ref.~\cite{BC} we finally obtain the following expression
for $C_4$
\begin{align}\label{C4-c} \notag
C_4& =\left(\frac{3 a_1}{80}+\frac{9 a_2}{160}+\frac{15 \zeta_5}{16} \right)\epsilon^{-2}
\\ \notag
&+ \left( -\frac{21 a_1}{80}-\frac{9 a_2}{80}+\frac{3 b_1}{80}+\frac{9 b_2}{160}+\frac{15 \zeta_3{}^2}{16}+\frac{5 \pi
   ^6}{2016} \right)\epsilon^{-1}
\\ \notag
&+ \bigg(\frac{741 a_1}{640}+\frac{807 a_2}{320}-\frac{21 b_1}{80}-\frac{9 b_2}{80}+\frac{3 c_1}{80}+\frac{9
   c_2}{160}-\frac{225 \zeta_7}{64}-\frac{5 \pi ^2 \zeta_5}{16}
\\
& \hspace*{50mm}   +\frac{7035 \zeta_5}{128}+\frac{81 \zeta_3{}^2}{16}+\frac{\pi ^4 \zeta_3}{32}-\frac{27 \zeta_3}{4}-\frac{237}{16} \bigg) + O(\epsilon)\,.
\end{align}
Here the constants $a_i, b_i$ and $c_i$ describe the contribution of the two non-planar
master integrals, Eq.~\p{master-np}. We recall that $C_4$ defines the five-loop correction
to the Konishi anomalous dimension and, therefore, it should be finite for $\epsilon\to 0$.
The condition for the $1/\epsilon^2$ and $1/\epsilon$ poles to cancel inside $C_4$ leads to two relations between the coefficients $a_i$ and $b_i$. As we shall see in a moment, these relations are indeed satisfied.

Let us start with the leading $O(1/\epsilon)$ term on the right-hand side of \p{master-np}.
The simplest way to compute the residue at the pole is to  Fourier transform the integral
into momentum  with the help of (\ref{Fourier}).
Notice that the expressions on the right-hand side of \p{master-np} are valid for $x_1^2=1$,
but their dependence on $x_1^2$ can easily be restored from dimension analysis. In this
way, we find from the first relation in  \p{master-np}
\begin{align}\label{F-I1}
{\cal F}\left[ I_1\right] = {\cal F}\left[\frac{a_1}{\epsilon}  (x_1^2)^{-4\epsilon}+O(\epsilon^0)\right] = \left(64\, a_1
+ O(\epsilon)\right)(p^2)^{-2+5\epsilon}\,.
\end{align}
This relation implies that the coefficient $a_1$ can be obtained from the Fourier
transformed integral ${\cal F}\left[ I_1\right]$ evaluated at $D=4$ dimensions. Transforming the
integral $I_1$ into the momentum representation  we find that
 ${\cal F}\left[ I_1\right]$  coincides (up to a factor of 16) with the conventional four-dimensional momentum Feynman integral denoted $N_0$ in Ref.~\cite{BC}
\begin{align}
 {\cal F}\left[ I_1\right] = 16 \left[ \parbox[c]{30mm}{\includegraphics[width=30mm]{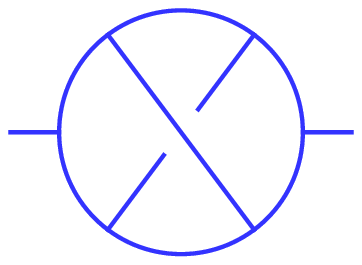}} \right]= 16\times\big(20\zeta_5 + O(\epsilon) \big) (p^2)^{-2+5\epsilon}\,.
\end{align}
Comparing this relation with \p{F-I1} we find
\begin{align}\label{a1}
a_1 =
5\zeta_5 \,.
\end{align}
Let us now turn to the integral $I_2$ in  \p{master-np} and Fourier transform it
\begin{align}\label{F-I2}
{\cal F}[I_2]=  F\left[\frac{a_2}{\epsilon}  (x_1^2)^{-1-4\epsilon}+O(\epsilon^0)\right] =4\left(\frac{a_2}{\epsilon} + O(\epsilon)\right) (p^2)^{-1+5\epsilon} \,.
\end{align}

To identify the momentum integral corresponding to ${\cal F}[I_2]$ we have to Fourier transform
all factors in the denominator of $I_2$ including $1/(x_{19}^2)^2$. In that case, we
find from \p{Fourier}
\begin{align}
{\cal F}\left[\frac1{(x_{19}^2)^2}\right] =2^{-2\epsilon} \Gamma(-\epsilon) (p^2)^\epsilon = -\frac1{\epsilon} + O(\epsilon^0)\,.
\end{align}
The fact that the residue at the pole in this relation does not depend
on the momentum $p$ implies that the corresponding line in the Feynman diagram
shrinks to a point. As a result,
\begin{align}
{\cal F}[I_2] =   -\frac4{\epsilon}\left[ \parbox[c]{30mm}{\includegraphics[width=30mm]{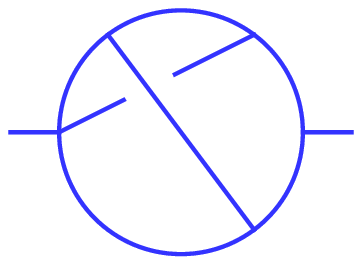}} \right] = -\frac4{\epsilon}\left[ \parbox[c]{30mm}{\includegraphics[width=30mm]{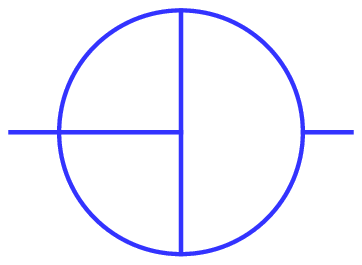}} \right] =  -\frac4{\epsilon}\times\lr{20\zeta_5 + O(\epsilon)} (p^2)^{-1+5\epsilon}\,.
\end{align}
Here in the second relation we redrew the same diagram, so that it takes the form of the
diagram $N_2$ in the notation of Ref.~\cite{BC}. Comparing the last relation with \p{F-I2}
we conclude that
\begin{align}\label{a2}
a_2=-20 \zeta_5\,.
\end{align}

Let us now compute the subleading terms in the expansion \p{master-np}. To this end,
we introduce the following auxiliary integrals (with $D=4-2\epsilon$)
\begin{align}\notag
I_3(\kappa) &=\frac{\e^{4\gamma  \epsilon}}{\pi^{2D}} \int\frac{d^{D}x_6 d^{D}x_7 d^{D}x_8 d^{D}x_9}{(x_{16}^2x_{19}^2 x_{67}^2x_{68}^2x_{78}^2x_{79}^2x_{7}^2x_{8}^2x_{89}^2)^{1-\epsilon \kappa}}  \,,
\\ \label{aux}
I_4(\kappa) &=\frac{\e^{4\gamma  \epsilon}}{\pi^{2D}} \int\frac{d^{D}x_6 d^{D}x_7 d^{D}x_8 d^{D}x_9}{(x_{16}^2x_{17}^2x_{19}^2 x_{67}^2x_{68}^2x_{79}^2x_{7}^2x_{8}^2x_{89}^2)^{1-\epsilon \kappa}} \,,
\end{align}
with $\kappa$ being a parameter. The diagrammatic representation for these integrals (for $\kappa=0$) is shown
in Fig.~\ref{fig-basis-np}. For finite $\kappa$ the two integrals are finite as $\epsilon\to 0$  and, therefore, they admit an expansion in powers of $\epsilon$ and $\kappa\epsilon$
\begin{align}\label{b-c}
I_i(\kappa) = b_i  + \epsilon \lr{ c_i + \kappa\, d_i}  + O(\epsilon^2) \,, \qquad (i=3,4)\,,
\end{align}
where we put $x_1^2=1$.
Notice that the leading $O(\epsilon^0)$ term does not depend on $\kappa$, whereas the $O(\epsilon)$ term is a linear of function of $\kappa$.

A distinguishing feature of the integrals \p{aux} is that, in the special case $\kappa=0$,
the IBP relations allow  us express $I_3(0)$ and $I_4(0)$  in terms of two master integrals $I_1$ and $I_2$, Eqs.~\p{master-np}. More precisely,
\begin{align}\label{system} \notag
b_3&= -\frac{2 }{3}\,b_1-\frac{7}{3}\,b_2 -70 \,\zeta_5
+\frac{26}{3} \zeta_3{}^2-\frac{65}{567} \pi ^6\,,
\\\notag
b_4& = -b_1-2 b_2-45 \zeta_5+7 \zeta_3{}^2-\frac{5}{54} \pi^6\,,
\\ \notag
c_3&=\frac{14}{3}\, b_1 +\frac{14
   }{3}\,b_2-\frac{2 }{3} \,c_1-\frac{7 }{3}\,c_2
 -\frac{4667}{6} \zeta_7+\frac{130}{9} \pi ^2 \zeta_5-\frac{100}{3}\zeta_5+\frac{13}{45} \pi ^4 \zeta_3\,,
\\
c_4& = 2b_1 -6 b_2-c_1-2
   c_2-\frac{4193}{4} \zeta_7+\frac{35 }{3}\pi ^2 \zeta_5-275
   \zeta_5+35 \zeta_3{}^2+\frac{7}{30} \pi ^4 \zeta_3-\frac{25}{54} \pi ^6\,.
\end{align}
Once the auxiliary integrals \p{b-c} have been computed, we could use these relations to obtain the needed coefficients $b_1,b_2$ and $c_1, c_2$.

One may wonder why we introduced the parameter $\kappa$ into the definition of the integrals \p{aux} if we only need its value at $\kappa=0$. The reason
for this is that, as we will see in a moment, it is much easier to compute the integrals \p{aux} for the two special values $\kappa=1/2$ and $\kappa=1$. Then,
taking into account that $I_i(\kappa)$ is a linear function of $\kappa$ at order $O(\epsilon)$, Eq.~\p{b-c},  we find
\begin{align}
 I_i(0) = 2 I_i(1) - I_i(1/2)+ O(\epsilon^2)  = b_i + \epsilon c_i+ O(\epsilon^2) \,.
\end{align}
In what follows, we shall evaluate $I_i(1)$ and $I_i(1/2)$ and, then, apply this relation to compute $b_3,b_4$ and $c_3,c_4$.

Let us consider the integrals \p{aux} for $\kappa=1$. In this case it is easy to see that the integrand is given by a product of factors $1/(x^2)^{1-\epsilon}$
which coincide with scalar propagators in $D=4-2\epsilon$ dimensions. As a result, upon the Fourier transform,
the integrals ${\cal F}[I_3(1)]$ and ${\cal F}[I_4(1)]$
are given by conventional four-loop momentum Feynman integrals. In this way, we find that the integral ${\cal F}[I_3(1)]$ coincides with the master integral
$M_{45}$ in the notation of \cite{BC}
\begin{align}\label{I3-res}
I_3(1) &= G_0^4M_{45} =  36 \zeta_3{}^2+\epsilon\lr{108\,\zeta_3\zeta_4+288 \zeta_3{}^2-378\,\zeta_7} + O(\epsilon^2)\,,
\end{align}
where the additional factor
$G_0=\e^{\gamma\epsilon}\Gamma(1+\epsilon)\Gamma(1-\epsilon)^2/\Gamma(2-2\epsilon)=1+2\epsilon+ O(\epsilon^2)$
is inserted to convert the result of Ref.~\cite{BC} obtained in the $G-$scheme to the regularization scheme used in \p{aux}.
The second momentum integral
${\cal F}[I_4(1)]$ is not a master integral.  We applied {\tt FIRE} to reduce it to the master integrals of Ref.~\cite{BC}
and arrived at the following result:
\begin{align}\notag
 I_4(1) =& - M_{01}\frac{(3-4 \epsilon) (1-4 \epsilon) (4-5 \epsilon) (3-5 \epsilon) \left(10-105
   \epsilon+326 \epsilon ^2-319 \epsilon ^3\right)}{6 (1-\epsilon) \epsilon ^5 (1-3 \epsilon)}
 \\ & \notag
   + M_{11} \frac{4 (1-2 \epsilon) (2-3 \epsilon) (1-3 \epsilon) (3-4 \epsilon) (1-4 \epsilon)}{3 (1-\epsilon) \epsilon ^4}
     -M_{35}\frac{2 (1-4 \epsilon) (1-5 \epsilon)}{3 (1-\epsilon) \epsilon }
 \\ &  \notag
   -M_{13}\frac{(1-2 \epsilon) (2-3 \epsilon) (3-5 \epsilon) (2-9 \epsilon) (7-19 \epsilon)}{6 (1-\epsilon) \epsilon ^4}
   -M_{36}\frac{1-5 \epsilon}{1-\epsilon}
 \\ &
   +M_{12}\frac{(1-2 \epsilon) (2-3 \epsilon)^2 (1-3 \epsilon)^2}{(1-\epsilon) \epsilon ^4}
   +M_{21}\frac{4 (1-2 \epsilon)^3 (1-4 \epsilon)}{3 (1-\epsilon) \epsilon ^3}\,.
\end{align}
Replacing the basis integrals by their explicit expressions we get
\begin{align}\label{I4-res}
 I_4(1) &= 36 \zeta_3{}^2 +   {\epsilon} \left( 108\,\zeta_3\zeta_4+108 \zeta_3{}^2+\frac{189}{2}\zeta_7\right)+O(\epsilon^2) \,,
\end{align}
where we put $x_1^2=1$.

Let us now examine the integrals \p{aux} for $\kappa=1/2$. In this case, the special feature of the integral $I_3(1/2)$ is that the conformal weight of the integrand at the integration points $x_7$ and $x_8$ equals the space-time dimension $4(1-\kappa\epsilon)=D$. As a consequence,
performing inversion $x_i^\mu \to x_i^\mu/x_i^2$ we obtain the following representation for $I_3(1/2)$ (at $x_1^2=1$)
\begin{align}\label{I3-half}
I_3(1/2) &=\frac{\e^{4\gamma  \epsilon}}{\pi^{2D}} \int\frac{d^{D}x_6 d^{D}x_7 d^{D}x_8 d^{D}x_9}{(x_{16}^2x_{19}^2 x_{67}^2x_{68}^2x_{78}^2x_{79}^2x_{6}^2x_{9}^2x_{89}^2)^{1-\epsilon/2}}  = \  \parbox[c]{30mm}{
\psfrag{x1}[cc][cc]{$\scriptstyle x_{1}$}
\psfrag{x5}[cc][cc]{$\scriptstyle 0$}
\psfrag{x6}[cc][cc]{$\scriptstyle x_{6}$}
\psfrag{x7}[cc][cc]{$\scriptstyle x_{7}$}
\psfrag{x8}[cc][cc]{$\scriptstyle x_{8}$}
\psfrag{x9}[cc][cc]{$\scriptstyle x_{9}$}
\includegraphics[width=30mm]{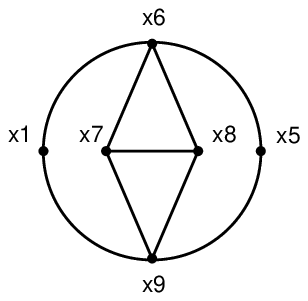}}  \,,
\end{align}
where on the right-hand side we depicted the corresponding Feynman diagram.
Compared with the first relation in \p{aux}, the product $x_7^2 x_8^2$ in the denominator gets replaced here with $x_6^2 x_9^2$.
We observe that the diagram on the right-hand side of \p{I3-half} contains a two-loop subgraph. As a consequence,
the integration over $x_7$ and $x_8$ can be easily performed with the help of the relation \cite{Chetyrkin:1980pr,MINCER}
\begin{align}\label{iden}
 \frac{\e^{2\gamma  \epsilon}}{\pi^{D}} \int\frac{d^{D}x_7 d^{D}x_8}{(x_{67}^2x_{68}^2x_{78}^2x_{79}^2x_{89}^2)^{1-\epsilon/2}} = \frac{1}{(x_{69}^2)^{1-\epsilon/2}}\big[6 \zeta_3 + (9 \zeta_4+12 \zeta_3) \epsilon + O(\epsilon^2)\big]\,.
\end{align}
Substituting this relation into \p{I3-half} we find that the remaining integral over $x_6$ and $x_9$ takes the same form as \p{iden} leading to
\begin{align}
 I_3(1/2) &= \big[6 \zeta_3 + (9 \zeta_4+12 \zeta_3) \epsilon + O(\epsilon^2)\big]^2 \,.
\end{align}
It remains to determine $I_4(1/2)$. The corresponding Feynman diagram has the same form as the one for $I_4(1/2)$ (see Fig.~\ref{fig-basis-np})
with the only difference that each solid line carries the index $(1-\epsilon/2)$. It is easy to see that the integrals $I_4(1/2)$ and $I_3(1/2)$ look
differently. Quite remarkably, as we will show later in this appendix, they coincide leading to
\begin{align}\label{I3I4-res}
I_4(1/2) = I_3(1/2) = 36\zeta_3{}^2 + \epsilon \left(108\zeta_3 \zeta_4 + 144 \zeta_3{}^2\right)+ O(\epsilon^2)\,.
\end{align}
We would like to stress that this relation is exact and it holds for arbitrary $\epsilon$.

Then, we combine the relations \p{I3-res}, \p{I4-res}, \p{I3I4-res} together and  find from \p{b-c}
\begin{align}\notag
I_3(\kappa)&=36  \zeta_3{}^2+\epsilon\lr{108\,\zeta_3 \zeta_4 +288\,\kappa \zeta_3{}^2+(1-2\kappa)378\,\zeta_7} + O(\epsilon^2)\,,
\\[2mm]
I_4(\kappa)&=  36 \zeta_3{}^2+\epsilon\lr{108\,\zeta_3 \zeta_4 +(180-72\kappa) \zeta_3{}^2-\ft{189}2\lr{1-2\kappa}\,\zeta_7} + O(\epsilon^2)\,.
\end{align}
Matching these expressions into \p{b-c} we obtain the following relations for the coefficients
\begin{align}\notag
b_3& =b_4= 36  \zeta_3{}^2\,,
\\[2mm] \notag
c_3&=108\,\zeta_3 \zeta_4 + 378\,\zeta_7\,,
\\
c_4&=108\,\zeta_3 \zeta_4 +180 \zeta_3{}^2-\frac{189}2 \,\zeta_7\,.
\end{align}
Their substitution into \p{system} yields a system of linear relations for the coefficients $b_1,b_2$ and $c_1,c_2$ whose solution is
\begin{align}\notag
a_1 & =5\,\zeta_5\,, && \hspace*{-80mm} b_1 = \frac{5}{378} \pi^6 - 13\, \zeta_3{}^2 +
  35 \,\zeta_5 \,,
\\[2mm] \notag
a_2 &=-20\,\zeta_5\,, && \hspace*{-80mm} b_2 = -\frac{10}{189} \pi^6 - 8\, \zeta_3{}^2 -
  40\, \zeta_5\,,
\\[2mm] \notag
c_1 &=-\frac{13}{30}\pi ^4 \zeta_3-91 \zeta_3{}^2+195 \zeta_5 -\frac{5 }{3}\pi ^2
   \zeta_5+\frac{345 }{4}\zeta_7+\frac{5 }{54} \pi ^6\,,
\\[2mm] \label{c12}
c_2 &= -\frac{4}{15}\pi ^4 \zeta_3-16 \zeta_3{}^2-80 \zeta_5+\frac{20 }{3}\pi ^2 \zeta_5-520 \zeta_7-\frac{20}{189} \pi ^6\,.
\end{align}
Substituting these relations in \p{C4-c} we verify the cancellation of poles in $\epsilon$ and reproduce \p{main}.

\begin{figure}[t]
\psfrag{x1}[cc][cc]{$\scriptstyle x_{1}$}
\psfrag{x5}[cc][cc]{$\scriptstyle 0$}
\centerline{\includegraphics[width=0.5\linewidth]{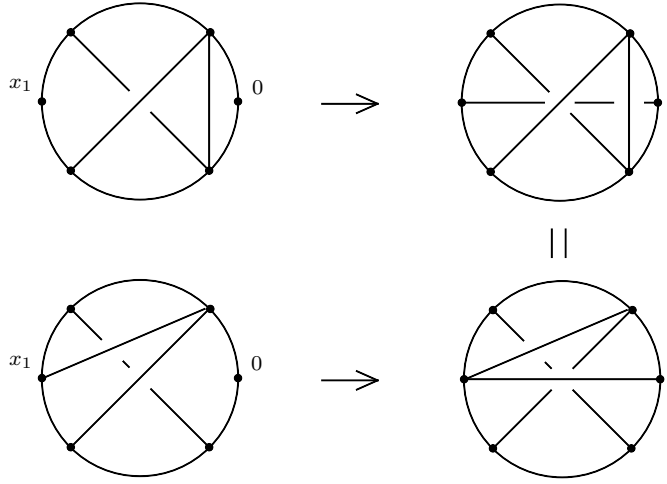}}
\caption{Glue procedure for the integrals $I_3$ and $I_4$ defined in \p{Ii-gen}.
All lines correspond to factors of $1/(x^2)^\alpha$ with the same index $\alpha=1-\epsilon/2-\lambda/10$. The dots with the labels $x_1$ and $0$ describe the external points, the remaining four
dots describe the integration points.}
\label{fig-glue}
\end{figure}

We complete this appendix with  a proof of the relation  $I_4(1/2) = I_3(1/2)$. It relies on applying the
cut-and-glue method of Ref.~\cite{IBP,BC}\footnote{It turns out that this generalized gluing
is very close in its spirit to the strategy of ref.~\cite{GorIs} where
Feynman integrals were considered as distributions with respect to the parameter of
analytic regularization.}.
Let us examine the integrals \p{aux} for $\kappa=1/2+\lambda/(10 \epsilon)$ with $\lambda$ arbitrary. Dimensional analysis shows that the integrals
have the following form
\begin{align}\label{Ii-gen}
I_i(1/2+\lambda/(10\epsilon)) =  \frac{c_i(\epsilon,\lambda)}{(x_1^2)^{1-\epsilon/2 - 9\lambda/10}} \qquad (i=3,4)\,,
\end{align}
with $c_i$ being some function of $\epsilon$ and $\lambda$. Then, the relation  $I_4(1/2) = I_3(1/2)$ implies that for arbitrary $\epsilon$
\begin{align}\label{magic}
c_3(\epsilon,0) = c_4(\epsilon,0)\,.
\end{align}
To show this, we consider the following Fourier integral
\begin{align}
{\cal F}\left[ \frac{ I_i(1/2+\lambda/(10\epsilon))}{(x_1^2)^{1-\epsilon/2 - \lambda/10}}\right] = {\cal F}\left[ \frac{c_i(\epsilon,\lambda)}{(x_1^2)^{2-\epsilon  - \lambda}}\right] =c_i(\epsilon,\lambda) \frac{2^{-2\lambda}\Gamma(\lambda)}{\Gamma(2-\epsilon-\lambda)} (p^2)^{-\lambda}\,,
\end{align}
where in the second relation we applied (\ref{Fourier}). A crucial observation is that for small $\lambda$ the expression on the right-hand side develops
a pole $1/\lambda$ with the residue independent on the momentum $p$
\begin{align}
{\cal F}\left[ \frac{ I_i(1/2 +\lambda/(10\epsilon))}{(x_1^2)^{1-\epsilon/2 - \lambda/10}}\right] =
 {\lambda^{-1}}\frac{c_i(\epsilon,0)}{\Gamma(2-\epsilon)}  + O(\lambda^0)\,.
\end{align}
The integrals $I_i(1/2 +\lambda/(10\epsilon))$ (with $i=3,4$) are described by the   Feynman diagrams shown in Fig.~\ref{fig-glue} in the left column. All solid lines in
these diagrams correspond to factors of $1/(x^2)^{1-\epsilon/2-\lambda/10}$.
Then, dividing $I_i$ by $(x_1^2)^{1-\epsilon/2 - \lambda/10}$ amounts to adding
one additional line to the diagram connecting the external points $x_1$ and $0$.
The resulting diagrams are shown in the right column of Fig.~\ref{fig-glue}.
Notice that, up to changing the labels of the dots, these two graphs coincide (to see this,
it suffices to rotate the lower diagram clockwise by $2\pi/3$). After taking the Fourier
transform, in the momentum representation, the momenta $p$ and $-p$
are injected into the external points $x_1$ and $0$, respectively,
while for the remaining integrated points  the corresponding
momentum equals zero. As was argued in \cite{BC}, the very fact that the leading
asymptotic behaviour of the Fourier integral for $\lambda\to 0$
does not depend on the momentum $p$ implies that the residue
at the pole $1/\lambda$ is not sensitive to the choice of the external points. This
is exactly what happens for the two graphs in the right column of Fig.~\ref{fig-glue},
the only difference between these graphs is in the assignment of the external
points. Since their residue at the pole $1/\lambda$ are defined by the functions
$c_3(\epsilon,0)$ and $c_4(\epsilon,0)$, we conclude that they are equal to each other
leading to \p{magic}.

To check numerically our analytic results for these two non-planar integrals we used the code {\tt FIESTA} \cite{FIESTA}
which gave the precision of six digits.

\section{Twist-two anomalous dimensions}
\label{sec:twist-two-anomalous}
In Section~\ref{sect:OPE}, we have applied the OPE \p{OPE0} to identify the contribution
of the Konishi operator to the four-point correlation function in the double-short distance $x_2\to x_1$ and $x_4\to x_3$. In this appendix, we address the
larger class of operators of twist two, of which the Konishi operator is the simplest (spin zero) representative.

For the {\it protected} scalar operators  \p{defO} the OPE takes
the following general form:
\begin{align}\label{OPE-gen}
{\mathcal O}(x_1,y_1) {\mathcal O}(x_2,y_2) = \sum_{\Delta, \, S,\, \mathcal{R}}   \frac{C^{\mathcal{R}}_{{\mathcal O} {\mathcal O} O_\Delta}(y_1,y_2)}{(x_{12}^2)^{2-\frac12(\Delta -S)}} (x_{12})_{\mu_1}\ldots (x_{12})_{\mu_S}\left[ O_\Delta^{\mu_1\ldots\mu_S;\mathcal{R}}(x_2)+\dots\right] .
\end{align}
Here the sum on the right-hand side runs over conformal primary operators
$O_\Delta^{\mu_1\ldots\mu_S;\mathcal{R}}$ carrying Lorentz spin $S$, scaling dimension
$\Delta$ and the dots denote the contribution of their conformal descendants.
The relation \p{OPE-gen} generalizes \p{OPE0} which describes the most singular contribution
of operators with the lowest value of $\Delta$. Also, since each operator
${\mathcal O}(x_i,y_i)$ belongs to the representation $\mathbf{20'}$ of the $R$ symmetry group $SU(4)$, the right-hand side of \p{OPE-gen} involves the sum over all irreducible
representations $\mathcal{R}$ that appear in the tensor product
\begin{align}\label{R}
\mathbf{20'}\times\mathbf{20'} = \mathbf{1} + \mathbf{15} + \mathbf{20'} + \mathbf{84} + \mathbf{105} +\mathbf{175} \,.
\end{align}
These representations can be identified from the $y-$dependence of the operators $ O_\Delta^{\mu_1\ldots\mu_S;\mathcal{R}}(x_2)$ in the expansion.  The contribution of each operator to the right-hand side of \p{OPE-gen}  is accompanied by the coefficient function $C^{\mathcal{R}}_{{\mathcal O} {\mathcal O} O_\Delta}$. It determines the three-point correlation function
$\vev{{\mathcal O}(1) {\mathcal O}(2) O_\Delta^{\mu_1\ldots\mu_S;\mathcal{R}}(x_3)}$
and depends, in general, on the coupling constant.

A notable example of the operators $O_\Delta^{\mu_1\ldots\mu_S;\mathcal{R}}$ that
shall play a special role in our discussion are the twist-two operator $\mathcal{O}_S$. They are $SU(4)$ singlet bilinear operators with arbitrary (even) Lorentz spin $S$ and
naive scaling dimension $\Delta^{(0)}=2+S$. The Konishi operator is the special
case of such operators with spin zero, $S=0$. The scaling dimensions of the
twist-two operators acquire an anomalous contribution:
\begin{align}\label{S-anom}
& \Delta_{S} \, = \, S+2 +
\gamma_S(a) = S+2 + \sum_{\ell=1}^\infty a^\ell \gamma_S^{(\ell)} \,,
\end{align}
with $ \gamma_S^{(\ell)}$ being non-trivial functions of $S$.

Applying \p{OPE-gen}, we find that every operator contributing to the OPE
gives a definite contribution to the four-point function \p{cor4loop} known as a conformal partial wave amplitude (or CPWA) \cite{FGG}
\begin{align}\label{cpw}
G(1,2,3,4)  = \frac{1}{x_{12}^4 x_{34}^4} \sum_{\Delta, \, S,\, \mathcal{R}} G_{\Delta, \, S}^{(\mathcal{R})}(u,v)\,,
\end{align}
where $G_{\Delta, \, S}^{(\mathcal{R})}(u,v)$ describes the contribution of the conformal
primary operator $O_\Delta^{\mu_1\ldots\mu_S;\mathcal{R}}$ and its conformal descendants. The conformal partial waves $G_{\Delta, \, S}^{(\mathcal{R})}(u,v)$
are definite functions of the conformal ratios $u$ and $v$ defined in \p{cr}. For
$u\to 0$ and $v\to 1$ they have the following asymptotic behaviour \cite{Dolan:2001tt}
\begin{align}\label{G-as}
G_{\Delta, \, S}^{(\mathcal{R})}(u,v) \sim u^{(\Delta-S)/2} (1-v)^S\left[ 1+ O(u,1-v)\right]\,.
\end{align}
Expanding the known result for the four-point correlation function over the CPWA  and applying \p{G-as}, we can extract the anomalous dimension of the operator.
For more information see, for example~\cite{Arutyunov:2001mh,Dolan:2001tt}.

In $\mathcal{N}=4$ SYM, the twist-two operators are the ground states (or superconformal primaries) of  ``long" (or unprotected) supermultiplets. Each state (or superdescendant) in such a multiplet has different naive conformal dimension, Lorentz spin and $SU(4)$ quantum numbers, but they all share the same anomalous  dimension  $\gamma_S(a)$. Therefore, to determine the anomalous dimension of the twist-two operators, we may look at the state in the multiplet which is most convenient to identify in the CPWA expansion \p{cpw}. As pointed out in  \cite{Dolan:2001tt}, the best choice is the twist-six state  corresponding to the $SU(4)$ channel $\mathbf{105}$ in \p{R}.  The advantage of this choice is that the twist-two supermultiplet has only one state in this $SU(4)$ channel, while for all other choices there are multiple candidates.\footnote{Choosing the state  in the $\mathbf{105}$ is helpful, but not indispensable for carrying out the CPWA analysis of the twist-two operators. The two-loop anomalous dimension of the $SU(4)$ singlet state of spin two was found for the first time in \cite{Arutyunov:2001mh}, using a form of the CPWA expansion different from that in  \cite{Dolan:2001tt}.  }

The specific correlation function which singles out the state in the $\mathbf{105}$ has the form
\begin{align}\label{Ot}
G_4(1,2,3,4) = \vev{ \cO(x_1) \cO(x_2) \tilde \cO(x_3) \tilde \cO(x_4) }\,,
\end{align}
where $\cO=\tr(ZZ)$, $\tilde \cO=\tr(\bar Z \bar Z)$  and $Z=\Phi^1+ i \Phi^2$ is a
complex scalar field. It can be obtained from the general expression
for the correlation function \p{cor4loop} by choosing the harmonic variables
as $Y_1=Y_2=(1,i,0,0,0,0)$ and $Y_3=Y_4=(1,-i,0,0,0,0)$.
Then, the relations  \p{cor4loop} and \p{intriLoops}  take the following form
\begin{align}\label{G-red}
G_4(1,2,3,4)  = \frac{2
\, (N_c^2-1)}{(4\pi^2)^{4}} \times {1 \over x_{13}^2 x_{24}^2} \times
{u \over v}  \times \sum_{\ell\ge 1} a^\ell  F^{(\ell)}(x_i)
\ ,
\end{align}
where the additional factor of $u/v$ comes from the function $R(1,2,3,4)$ (see Ref.~\cite{Eden:2011we}).
Then, the CPWA expansion
\p{cpw} of the correlation function \p{Ot} is (for  $u\rightarrow 0$)\footnote{Furthermore, by
considering only the limit $u\rightarrow 0$ with $v$ finite, we are
restricted to the sector where only the  twist-two operators together with their superdescendants
survive. Here it is known that there is a
unique twist-two, spin $S$ supermultiplet for each even $S$.}
\begin{align}\notag
G_4(1,2,3,4)
  \sim &\frac{ N_c^2-1 }{(4\pi^2)^{4} }
  {1 \over (x_{12}^2 x_{34}^2)^2}  \sum_{S/2 \in \mathbb{Z}^+}   A_S(a)\,
  u^{3+\gamma_S(a)/2} \, \bar v^S
\\ \label{G-cpw}
&\times
  \,{}_2F_1\left(3+S+\half\gamma_S(a),3+S+\half\gamma_S(a),6+2S+\gamma_S(a)|\bar v\right) \ .
\end{align}
where $\bar v=1-v$ and  the sum runs over non-negative even spins $S$ and
  the coefficient function $A_S(a)$ and the anomalous dimension
$\gamma_S(a)$ are functions of the 't Hooft coupling $a$
\begin{align}\label{C-coef}
A_S(a)=\sum_{\ell\ge 1} a^\ell A_S^{(\ell)}\,,\qquad \gamma_S(a)=\sum_{\ell\ge 1} a^\ell \gamma_S^{(\ell)}\,.
\end{align}
Comparing \p{G-red} and \p{G-cpw} we thus obtain (for $u\to 0$)
\begin{align}\label{eq:9}
{2  x_{13}^2 x_{24}^2\sum_{\ell\ge 1} a^\ell F^{(\ell)} } = \sum_{S/2 \in \mathbb{Z}^+} A_S\,
  u^{\gamma_S/2}\, (1-\bar v)\,\bar v^S
  \,{}_2F_1\left(3+S+\half\gamma_S,3+S+\half\gamma_S,6+2S+\gamma_S|\bar v\right) \,.
\end{align}
To make use of this relation, we expand both sides of this relation in powers of the
coupling constant $a$ and compare the coefficients of the powers of $\ln u$.
 Furthermore, on the
right-hand side, the contribution of the operator with spin $S$ is suppressed by a factor of
$\bar v^S$. Therefore, expanding both sides of \p{eq:9} for $\bar v\to 0$ up to
$O(\bar v^S)$ terms and equating the coefficients of the powers of $\ln u$ and $\bar v$,
we obtain equations for the expansion coefficients of $A_{S'}(a)$ and $\gamma_{S'}(a)$ for  $S'\le S$.

The main goal of this appendix is to apply \p{eq:9} to extract the coefficients \p{C-coef} at three loops.
To this end,    we have to supplement \p{eq:9} with the expressions for the correlation function $F^{(\ell)}$
for $\ell=1,2,3$ in the OPE limit.
At one and two loops, the corresponding expressions for  $F^{(1)}$ and $F^{(2)}$, Eq.~\p{F12}, are
known analytically \cite{davussladder}. At three loops, we can use the result for $F^{(3)}$ recently found in Ref.~\cite{Eden:2011we} (and described in
section~\ref{sec:three-loops}).
The explicit expression for $F^{(3)}$ in terms
of the basis scalar integrals is
\begin{align}
F^{(3)} = & 2 \, g(1,2,3,4) \, \left[ x^2_{12} x^2_{34} \, h(1,2;3,4) \, + \,
x^2_{13} x^2_{24} \, h(1,3;2,4) \, + \, x^2_{14} x^2_{23} \, h(1,4;2,3) \right]
\nonumber \\[2mm]
& +  6 \left[ L(1,2;3,4) \, + \, L(1,3;2,4) \, + \, L(1,4;2,3) \right]
\nonumber \\[2mm]
& +  4 \left[ E(1;3,4;2) \, + \, E(1;2,4;3) \, + \, E(1;2,3;4) \right]
 \nonumber \\[2mm] \label{loop3F}
& +  (1+1/v) \, H(1,2;3,4) \, + \, (1+u/v) \, H(1,3;2,4) \, + \,
 (1+u)\, H(1,4;2,3) \, .
\end{align}
Here the integrals $g,h$ were defined in \p{eq:g+h} while the remaining integrals $L, E$ and $H$ can be found in \cite{Eden:2011we}. The expressions for
the integrals $g,h,L$ are known in a closed analytic form. The
integrals $E$ and $H$ on the other hand are not currently known analytically. However, for our purposes
we only need their asymptotic behaviour at small $u$ and the first few terms of their
expansion in powers of $\bar v$.

The limit $u\to 0, \bar v\to 0$ can be described as the limit $x_1\to x_2$. For the $E-$ and $H-$integrals, we apply the well-known formulae \cite{AE}
for their asymptotic expansion in the limit $x_1\to x_2$ typical of Euclidean space
written in terms of a sum over certain subgraphs of a given
graph\footnote{
To find all the contributions to the asymptotic expansion in an automatic way we prefer to
use the code {\tt asy.m} \cite{asy} which reveals the contributions in the language of
regions. Observe that this code works not only for limits typical of Euclidean space but
also for other limits, in particular, for limits of Sudakov type.}.
Making use of conformal invariance, we set $x_4\to \infty$, $x_1=0$ and
arrive at the problem of analyzing the asymptotic behaviour of the Feynman integrals
depending on the two external coordinates $x_2$ and $x_3$ in the limit $x_2\to 0$.
The conformal ratios take the form
$u=x_{2}^2/x_{3}^2$ and $v=x_{23}^2/x_3^2$ so that for $x_2\to \rho x_2$ with
$\rho\to 0$ they scale in the Euclidean space as $u=O(\rho^2)$ and $\bar v=O(\rho)$.
We have evaluated terms up to order $O(\rho^4)$ for all integrals
entering the right-hand side of \p{loop3F} and obtained the following results
for the $E-$integrals
\begin{align}
x_{13}^2 x_{24}^2 E(1,3;2,4)
&=\ln u
\left[
  \lr{\ft{3 }{800}\zeta _3+\ft{17833}{5529600}}{\bar v}^4
+\lr{\ft{3}{512}\zeta _3+\ft{19}{6144}}  {\bar v}^3
+\lr{\ft{1}{96}\zeta _3+\ft{11}{4608}}{\bar v}^2
+\ft{3}{128}\zeta _3{\bar v}+\ft{3}{32} \zeta_3\right]
\nonumber\\&
+\lr{\ft{1261}{576000} \zeta_3-\ft{1060073 }{82944000}}{\bar v}^4
+\lr{\ft{5}{9216}\zeta_3-\ft{1423}{110592}}{\bar v}^3
-\lr{\ft{5}{1152}\zeta_3+\ft{37}{3456}}{\bar v}^2
-\ft{3 }{128}\zeta _3{\bar v}
-\ft{3}{16} \zeta _3,
\nonumber\\[10pt]
 x_{13}^2 x_{24}^2 E(1,4;2,3) 
&= \ \ln u
\left[
\lr{\ft{137}{3200} \zeta _3+\ft{28633}{5529600}}{\bar v}^4
+\lr{\ft{25}{512} \zeta _3+\ft{25}{6144}} {\bar v}^3
+\lr{\ft{11}{192} \zeta _3+\ft{11}{4608}} {\bar v}^2
+\ft{9}{128} \zeta _3{\bar v}
+\ft{3}{32} \zeta _3\right]
\nonumber\\&
+\lr{\ft{659}{144000} \zeta _3-\ft{1682573}{82944000}} {\bar v}^4
-\lr{\ft{65}{9216} \zeta _3+\ft{1865}{110592}} {\bar v}^3
-\lr{\ft{ 1}{36} \zeta _3+\ft{37}{3456}}{\bar v}^2
-\ft{9}{128}  \zeta _3{\bar v}
-\ft{3}{16} \zeta _3,
\nonumber\\[10pt]
  x_{13}^2 x_{24}^2 E(1,2;3,4)
 &= \left(
 \ft{2419}{2073600}{\bar v}^4
 +\ft{13 }{9216}{\bar v}^3
 +\ft{25}{13824}{\bar v}^2
 +\ft{1}{384}{\bar v}
 +\ft{1}{192}\right) (\ln u)^3
\nonumber\\&
 +\left(-\ft{24679}{4608000} {\bar v}^4
 -\ft{125}{18432}{\bar v}^3
 -\ft{259 }{27648}{\bar v}^2-\ft{1}{64}{\bar v}-\ft{3}{64}\right)(\ln u)^2
\nonumber\\&
+\left(\ft{10888367 }{1244160000}{\bar v}^4
+\ft{209}{18432}{\bar v}^3
+\ft{1361}{82944} {\bar v}^2
+\ft{1}{32}{\bar v}+\ft{5}{32}\right) \ln u
\nonumber\\&
+\lr{-\ft{1}{16}\zeta_5+\ft{29383}{1728000} \zeta_3+\ft{28271621}{7464960000}} {\bar v}^4
+\lr{-\ft{5}{64} \zeta_5+\ft{101}{4608} \zeta _3+\ft{1381}{331776}}{\bar v}^3
\nonumber\\&
+\lr{-\ft{5}{48} \zeta_5+\ft{209}{6912} \zeta_3+\ft{235}{55296}} {\bar v}^2
+\lr{-\ft{5}{32} \zeta_5+\ft{3}{64}\zeta _3} {\bar v}
-\ft{5}{16} \zeta_5+\ft{3}{32} \zeta_3-\ft{5}{32},
\end{align}
and for the $H-$integrals
\begin{align}
 \nonumber 
 x_{13}^2 x_{24}^2 H(1,3;2,4)
& =
\left(-\ft{1997}{6912000}{\bar v}^4-\ft{29}{110592}{\bar v}^3-\ft{1}{10368}{\bar v}^2+\ft{1}{1536}{\bar v}+\ft{1}{192}\right)
(\ln u)^3
\nonumber\\&
+\left(\ft{51643}{11520000} {\bar v}^4+\ft{575
}{110592}{\bar v}^3+\ft{217}{41472} {\bar v}^2-\ft{1}{16}\right) (\ln u)^2
\nonumber\\&
+\left(-\ft{25959283}{1244160000}{\bar v}^4-\ft{36013}{1327104} {\bar v}^3-\ft{8623
}{248832}{\bar v}^2-\ft{15
   }{512}{\bar v}+\ft{9}{32}\right) \ln u
\nonumber\\&
 +\lr{   -\ft{1997}{144000} \zeta_3+\ft{380181271}{12441600000}} {\bar v}^4
+\lr{-\ft{29}{2304} \zeta _3+\ft{344443}{7962624}} {\bar v}^3
\nonumber\\&
+\lr{-\ft{1}{216}\zeta_3+\ft{48113}{746496}}{\bar v}^2
+\lr{\ft{1}{32}\zeta_3+\ft{45}{512}}{\bar v}
   +\ft{1}{4}\zeta _3-\ft{7}{16},
\nonumber\\[10pt]
x_{13}^2 x_{24}^2 H(1,4;2,3) 
& =
\left(\ft{19253}{6912000} {\bar v}^4+\ft{119}{36864} {\bar v}^3+\ft{79
}{20736}{\bar v}^2+\ft{7}{1536} {\bar v}+\ft{1}{192}\right) (\ln u)
   ^3\nonumber\\&+\left(-\ft{1628267 }{69120000}{\bar v}^4-\ft{3155
}{110592}{\bar v}^3-\ft{371}{10368} {\bar v}^2-\ft{3
  }{64} {\bar v}-\ft{1}{16}\right) (\ln u)^2\nonumber\\&
  +\left(\ft{103801967}{1244160000} {\bar v}^4+\ft{136573
}{1327104}{\bar v}^3+\ft{33173
}{248832}   {\bar v}^2+\ft{95}{512} {\bar v}+\ft{9}{32}\right) \ln u
\nonumber\\&
+\lr{\ft{19253}{144000}\zeta _3-\ft{282347651}{3110400000}} {\bar v}^4
+\lr{\ft{119 }{768}\zeta _3-\ft{928859}{7962624}} {\bar v}^3
\nonumber\\&
+\lr{\ft{79}{432} \zeta _3-\ft{119557}{746496}} {\bar v}^2
 +\lr{\ft{7}{32} \zeta _3-\ft{125}{512}} {\bar v}+\ft{1}{4}\zeta
_3-\ft{7}{16}, \nonumber\\[10pt]
x_{13}^2 x_{24}^2 H(1,2;3,4) 
&=\ln u \left[
\lr{-\ft{13}{960}\zeta _3+\ft{149}{138240}}{\bar v}^4
+\lr{-\ft{1}{64}\zeta_3+\ft{1}{768}}{\bar v}^3
+\lr{-\ft{1}{64} \zeta_3+\ft{1}{768}}{\bar v}^2
   +\ft{3 }{16}\zeta
_3\right]
\nonumber\\&
+\lr{\ft{217}{11520} \zeta _3-\ft{791}{276480}} {\bar v}^4
+\lr{\ft{1}{32}
   \zeta _3-\ft{1}{256}}{\bar v}^3
+\lr{\ft{7}{128}\zeta _3-\ft{7}{1536}} {\bar v}^2 +\ft{3}{32} \zeta _3 {\bar v}-\ft{3}{16} \zeta
   _3.
\end{align}\label{eq:4}

Substituting these expressions into \p{eq:9} and going through the steps described above
we obtain the following results for the three-loop anomalous dimensions $\gamma_S(a)$ (for $S=0,2,4$)
\begin{align}\label{twi2} \notag
\gamma_0(a)&=3 a -3 a^2+\frac{21}{4} a^3+O\left(a^4\right),
\\ \notag
\gamma_2(a)&=\frac{25}{6}a -\frac{925 }{216}a^2+ \frac{241325}{31104}a^3 +O\left(a^4\right),
\\
\gamma_4(a)&=\frac{49}{10}a-\frac{45619}{9000}a^2
+\frac{300642097}{32400000} a^3+O\left(a^4\right),
\end{align}
and for the corresponding coefficients $A_S(a)$
\begin{align}\notag
 A_0(a) &=-a+a^2 \left(\frac{3 \zeta_3}{2}+\frac{7}{2}\right)-a^3 \left(2 \zeta_3+\frac{25 \zeta_5}{4} +12\right)+O\left(a^4\right),
\\ \notag
A_2(a) &=-a\frac{205}{1764}+a^2 \left(\frac{5 \zeta_3}{28}+\frac{76393}{148176}\right)-a^3 \left(\frac{1315 \zeta_3}{5292}+\frac{125 \zeta_5}{168}+\frac{242613655}{112021056}\right)+O\left(a^4\right),
\\ \notag
A_4(a) & =-a\frac{553}{54450}+a^2 \left(\frac{7 \zeta_3}{440}+\frac{880821373}{17249760000}\right)
\\   & \hspace*{40mm}
-a^3 \left(\frac{520093 \zeta_3}{26136000}+\frac{35 \zeta_5}{528}+\frac{1364275757197}{5692420800000}\right)+O\left(a^4\right).
\end{align}
We verified that the  relations \p{twi2} are in agreement with the results of \cite{Kotikov:2004er}.
Notice that $\gamma_0(a)$ coincides with the anomalous dimension of the Konishi
operator $\gamma_{\mathcal{K}}(a)$, Eq.~\p{Main}, and the expansion coefficients of $A_0(a)
=\sum_{\ell\ge 1} \alpha^{(\ell)} a^\ell$ coincide with the coefficients
$\alpha^{(\ell)}$ in \p{ex}.

Finally, we should  note that it is possible to invert this entire process,
namely to use predictions for the normalization and anomalous
dimensions of operators, and plug them into the conformal
partial wave expansion (\ref{cpw}) to obtain predictions for the
four-point correlation function.
More specifically, by using the predicted three-loop, all-spin, twist-two anomalous dimensions
  of~\cite{Kotikov:2004er}, together with two-loop and one-loop twist-two data, and plugging
  into~\p{G-cpw}, one can obtain the
three-loop correlation function in the limit $u\rightarrow 0$ with $v$ finite  (all
except the finite part as $u\rightarrow 0$ which would require three-loop normalization).
Summing
  the resulting expansion, one obtains a
closed analytic form for the correlation function in this limit as a sum of harmonic polylogarithms with argument
$\bar v$~\cite{hughfrancispaul}.

\end{document}